\documentclass[twocolumn]{aastex63}
\usepackage{nopageno}

\submitjournal{ApJ}

\shorttitle{Fujita et al.}
\shortauthors{Fujita et al.}

\begin{document}

\title{Origin of Weak \ion{Mg}{2} and Higher Ionization
  Absorption Lines in an Outflow from an Intermediate-Redshift Dwarf Satellite Galaxy}

\correspondingauthor{Akimi Fujita}
\email{fujitaa@shinshu-u.ac.jp}

\author[0000-0001-5091-6576]{Akimi Fujita}
\affiliation{Faculty of Engineering, Shinshu University,
4-17-1 Wakasato,
Nagano, Nagano 380-0926, Japan}

\author[0000-0002-5464-9943]{Toru Misawa}
\affiliation{School of General Studies, Shinshu University,
3-1-1, Asahi, Matsumoto City 390-8621, Japan}

\author[0000-0003-4877-9116]{Jane C. Charlton}
\affiliation{Department of Astronomy \& Astrophysics, 
The Pennsylvania State University, 
University Park, PA, 16802, USA}

\author[0000-0002-5451-9057]{Avery Meiksin}
\affiliation{SUPA\footnote{Scottish Universities Physics Alliance}, Institute for Astronomy, University of Edinburgh,
Blackford Hill, Edinburgh EH9 3HJ, UK}

\author[0000-0003-0064-4060]{Mordecai-Mark Mac Low} 
\affiliation{Department of Astrophysics,
  American Museum of Natural History, New York, NY 10024, USA}
\affiliation{Center for Computational Astrophysics, Flatiron
  Institute, New York, NY 10010, USA}




\begin{abstract}
Observations {at intermediate redshifts} reveal the presence of numerous, compact, weak Mg{\sc ii}
absorbers with near to super-solar metallicities, often surrounded by
more extended regions that produce C{\sc iv} and/or O{\sc vi} absorption in the
circumgalactic medium at large impact parameters from luminous
galaxies.  Their origin and nature remains unclear.
We hypothesize that undetected, satellite dwarf galaxies are
responsible for producing {some of} these weak Mg{\sc ii} absorbers. We test our
hypothesis using gas dynamical simulations of galactic outflows from a
dwarf satellite galaxy with a halo mass of
$5\times10^{9}$~M$_{\odot}$, which
     could form
in a larger $L^{*}$ halo at $z=2$, to {study the gas interaction in the
  halo.
We find that thin, filamentary, weak Mg{\sc ii} absorbers are produced in
two {stages}: 1) when shocked core collapse supernova (SNII) enriched
gas
     descending in a galactic fountain gets shock compressed by upward flows
     driven by subsequent SNIIs and cools
(\textit{phase 1}), and later, 
2) during an outflow driven by Type Ia supernovae {that} shocks and
sweeps {up} pervasive SNII enriched gas, which then cools
(\textit{phase 2})}. The width of the filaments and fragments are
$\lesssim~100$~pc,  and the smallest ones cannot be resolved at
12.8~pc resolution. The Mg{\sc ii} absorbers {in our simulations} 
are continuously generated for $>150$ Myr by shocks and cooling, though each
cloud survives for only $\sim60$~Myr. Their metallicity is 
10--20\% solar metallicity and column density is $<10^{12}$ cm$^{-2}$. 
They are also surrounded by larger (0.5--1 kpc) C{\sc iv} absorbers that seem to survive longer. In addition, larger-scale ($>1$~kpc) C{\sc iv} and O{\sc vi} clouds are produced in both expanding and shocked SNII enriched gas
which is photoionized by the UV metagalactic radiation at intermediate redshift.
Our simulation highlights the possibility of {dwarf} galactic outflows producing {highly enriched} multiphase gas.
\end{abstract}

\section{Introduction} \label{sec:intro}
Galactic outflows appear to regulate the structure and evolution of galaxies,
as they heat, ionize, and chemically enrich the surrounding circumgalactic medium (CGM) and even drive unbound winds that can reach the intergalactic medium (IGM) \citep[see e.g.][for reviews]{2015ARAA..53...51S, Heckman2017rev}.
A robust understanding of the stellar feedback processes driving these outflows, however, remains elusive. The observed properties of the outflows and outflow-CGM interaction at multiple wavelengths must be used to constrain theoretical models of the physics governing the outflows and outflow-CGM interaction. The most prominent observed properties are metal absorption lines, seen in the spectra of background quasars, that are believed to arise from inhomogeneities in the CGM. Numerical simulations are required to predict and interpret the observational signatures of these systems 
\citep[e.g.][]{Oppenheimer2012, Suresh2015, Keating2016, Turner2017, Oppenheimer18, Peeples2019}.

The derived metallicities of weak, low ionization absorbers are almost always greater than 10$\%$ solar and are often as high or even higher than the solar value. Some of them are even iron-enhanced compared with solar  \citep{Rigby2002, Charlton2003, Narayanan2008, Misawa2008, Lynch2007}. In addition, analyses of low-redshift absorbers show that there are fewer absorbers at present than in the past, and that most absorbers seem to live in group environments \citep{Muzahid2017}. With all the measured properties above, it is plausible to speculate that weak absorbers are created by transient processes, such as galactic outflows that carry metals and are less active in the modern Universe. {The outflows may} originate in satellite dwarf galaxies hosted by a larger halo that are too dim to be observed. 

The covering fraction of the weak absorbers is estimated to be $\gtrsim30\%$ in the CGM of galaxies brighter than 0.001L$^{*}$\citep{Narayanan2008,  Muzahid2017}. There would be on the order of a million tiny, weak absorbers {per galaxy} if a spherical geometry were assumed \citep{Rigby2002}. {It has been argued, however, that} weak absorbers reside instead in filamentary and sheet-like structures \citep{Milutinovic2006}.

Many of these systems show absorption by multiple high ionization species at the same velocity, often with additional components offset by 5-150 km s$^{-1}$ \citep{Milutinovic2006}. 
C{\sc iv} surveys at z~$\approx2\sim3$ in the environments of sub-Lyman Limit Systems (sLLS) suggest that C{\sc iv} clouds are more diffuse (n$_{HI}\sim10^{-4}$ to $10^{-3}$ cm$^{-3}$) and larger than Mg{\sc ii} clouds, with sizes between 0.1~kpc and 10~kpc \citep{Simcoe2004, Shaye2007, Lerner2016}. Some of C{\sc iv} clouds may have expanded from denser, more compact Mg{\sc ii} clouds \citep{Shaye2007}. These C{\sc iv} systems may be interpreted as being in photoionization equilibrium  at T$\sim10^{4}~K$, and their metallicities are found to be $\sim1\%$ solar to even solar or more \citep{Simcoe2004, Shaye2007, Lerner2016}. {There are also many O{\sc vi} absorption systems, which are more likely to have an origin in photoionized gas (rather than collisionally ionized gas) at z$\sim$2 due to the greater intensity of the EBR.} The detections of O{\sc vi} by \citet{Turner2014, Turner2015}, however, suggest the presence of a collisionally ionized gas phase for impact parameters $\lesssim$100 proper kpc  (pkpc) of large, star-forming galaxies at z$\sim2.4$.

In this paper, we test our hypothesis that 
galactic outflows from satellite dwarf galaxies, too dim to detect in the halo of a larger L$^{*}$ galaxy, produce compact weak Mg{\sc ii} absorbers surrounded by larger regions that produce C{\sc iv} and O{\sc vi} absorption. Using a small-scale hydrodynamical simulation of a dwarf galaxy, we find such structures are produced by repeated shocks and radiative cooling in the gaseous halo of the galaxy. We will highlight important physical processes at work which regulate the production of low and high ionization clouds, to be explored in larger-scale simulations in the next paper. 
We describe our numerical method in Section \ref{sec:hydro} and the dynamics of SNII and SNIa driven outflows and their interaction with surrounding gas, including the production of dense clumps and filaments, in Section \ref{sec:results}. In Section  \ref{sec:metals}, we study the distributions of weak Mg{\sc ii} absorbers and surrounding C{\sc iv} and O{\sc vi} absorbers in our simulation, and compare them to the properties of observed systems, followed by a resolution study (Section \ref{sec:resolution}) and a summary (Section \ref{sec:summary}). 

\section{Numerical Method} \label{sec:hydro}
We use the adaptive mesh refinement hydrodynamics code Enzo
\citep[]{Bryanetal2014} to simulate repeated supernova explosions in
the disk of a dwarf galaxy. We solve the equations of hydrodynamics using a direct-Eulerian piecewise parabolic method  \citep[]{Collela1984, Bryanetal2014} and a two-shock approximate Riemann solver with progressive fallback to more diffusive Riemann solvers in the event that higher order methods produce negative densities or energies. 
Our simulation box has dimensions ($L_x, L_y, L_z$) = (6.5536,
6.5536, 32.768)~kpc, initially with (32,  32, 160) cells. Only
half the galactic disk above its midplane is simulated. We refine
cells to resolve shocks with a standard minimum pressure jump
condition \citep[]{Collela1984} and to resolve cooling at turbulent
interfaces where the sound crossing time exceeds the cooling time. We use 4
refinement levels resulting in a highest resolution of 12.8~pc
(\textit{standard simulation}). We also ran the same simulation with 3
refinement levels as a comparative resolution study (\textit{low-res
  simulation}), and by applying 6 refinement levels in a region where
Mg{\sc ii} filaments form in order to test the effects of resolution on
fragmentation (\textit{high-res zoom simulation}). We assume a flat $\Lambda$CDM cosmology with the 2018 Planck Collaboration measured parameters $\Omega_{m}=0.315$, $\Omega_{\Lambda}=0.685$, $h=0.674$, and $\Omega_{b}=0.0493$ \citep{Aghanim2019}.

\subsection{Galaxy Model} \label{sec:galaxy}
We model a dwarf galaxy at redshift $z=2$ with a halo mass
$M_{\rm halo}=4\times10^{9}$~M$_{\odot}$, and a virial radius $R_{\rm vir}=17.3$\,kpc.
This model has a disk gas mass, $M_{\rm g}=5.2\times10^{8}$~M$_{\odot}$.
We adopt a \citet{Burkert1995} dark matter potential with a core radius $r_{0}=848$\,pc and central density $\rho_{0}=1.93\times10^{-23}$\,g~cm$^{-3}$, although this potential profile is a fit to the observed rotation curves of nearby dwarf galaxies rather than those at z=2. Our choice of $r_{0}$ and $\rho_{0}$ ensures that the resulting potential profile reproduces
 a \citet{NFW} dark matter potential with $c=12.2$ for the same dwarf halo at $r>400$~pc.
The gas is described as a softened exponential disk:
\begin{equation}
\rho(R,z) = \frac{M_{\rm g}}{2\pi a^2_{\rm g}b_{\rm g}} 0.5^2
\mathrm{sech}\left(\frac{R}{a_{\rm g}}\right)
\mathrm{sech}\left(\frac{z}{b_{\rm g}}\right)
\end{equation}
where $M_{\rm g}$ is the total mass of gas in the disk, and $a_{\rm
  g}$ and $b_{\rm g}$ are the radial and vertical gas disk scale
heights \citep{Tonnesen2009}. We chose $a_{\rm g}=621$\,pc based on the
exponential disk approximation of \citet{Mo1998}, with $\lambda=0.05$,
and $b_{\rm g}=160$\,pc based on the thin disk approximation
\citep[]{Toomre1963} with an effective sound speed, $c_{\rm s, eff}=11.3$~km~s$^{-1}$ \citep[]{Fujita2009}. Given this gas density distribution in the disk, the gas temperature and pressure are calculated to maintain the disk in hydrostatic equilibrium
with the surrounding halo potential in the $z$-direction, and the rotational velocity of the gas disk is set to balance the radial gravitational
force and the pressure gradient. The disk temperature varies between
$10^{3}$~K and a few $\times10^{4}$~K, and the maximum circular
velocity is $v_{\rm max}=48.8$~km~s$^{-1}$ with the escape velocity from
the potential $v_{\rm esc}=69.0$~km~s$^{-1}$. Our model galaxy is placed
in a static halo background with $ \rho_{\rm bg}=1.83
\times10^{-28}$~g~cm$^{-3}$ so that the gas mass within the virial radius is
M$_{\rm halo}\left(\frac{\Omega_{\rm b}}{\Omega_{\rm m}}\right)$. The metallicity of all the gas in the box is initially set at $Z=0.001$ with mean molecular weight $\mu=0.6$.
\vskip 0.2in
\subsection{Cooling} \label{sec:cooling}

\begin{figure}[h]
\includegraphics[width=\columnwidth]{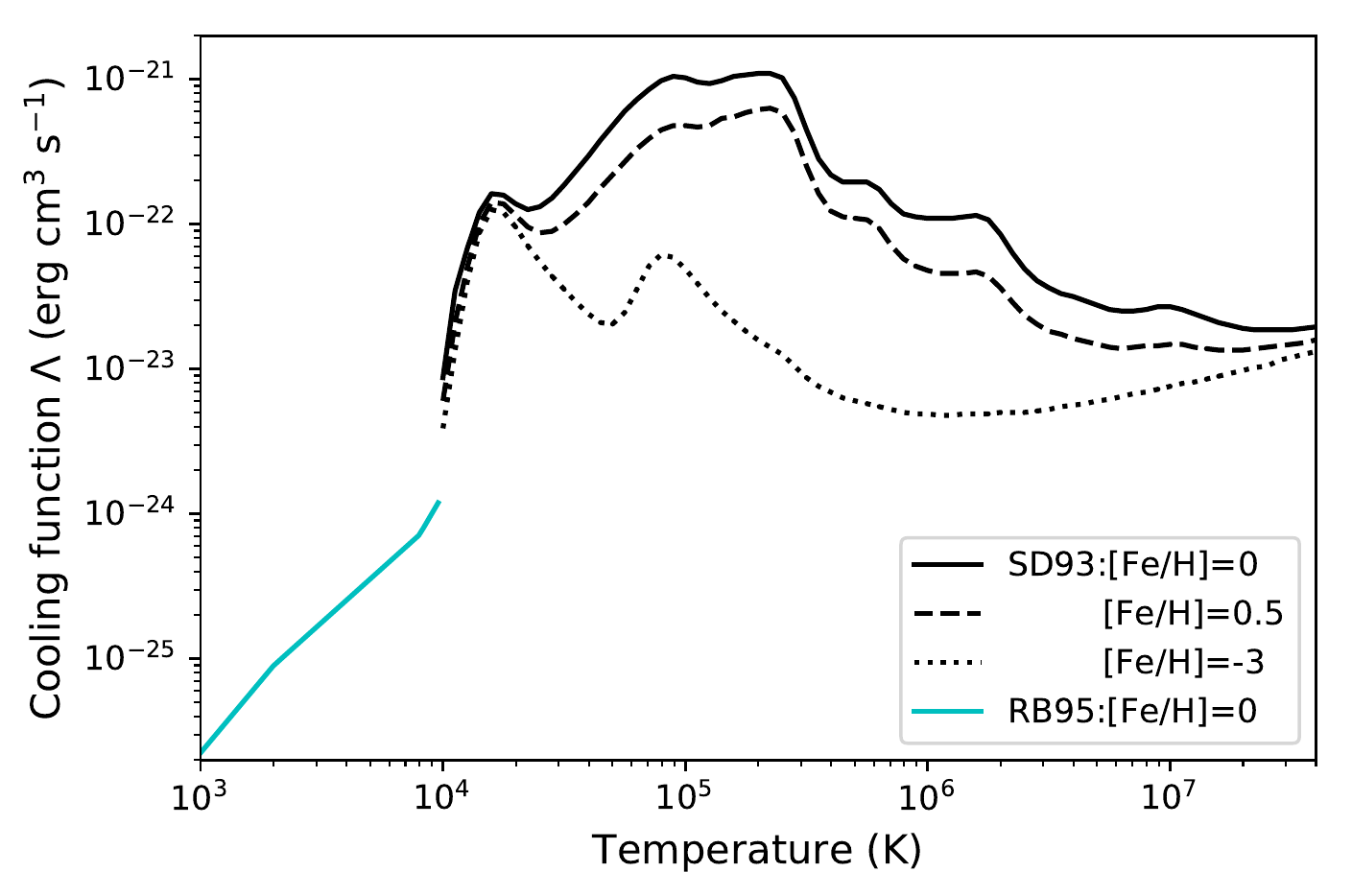}
\caption{Radiative cooling functions used in our simulations as a
  function of temperature $T$ from \citet[]{SD93} for $T\ge10^{4}$~K  for
  different metallicities and from \citet[]{Rosen1995} for $T < 10^{4}$~K for solar metallicity. \label{fig:cf}}
\end{figure}

Figure~\ref{fig:cf} shows the cooling curves used in our simulations. We use radiative cooling curves as a function of temperature above $10^{4}$~K for gas in collisional ionization equilibrium (CIE) with various metallicities: [Fe/H]=-3, -2, -1.5, -1, -0.5, 0, +0.5 \citep{SD93}.  A radiative cooling rate for gas in a cell with a metallicity is computed by interpolating between the cooling curves. Cooling of gas below temperature $10^{4}$~K is approximated with the cooling curve of \citet[]{Rosen1995} computed for solar metallicity. Although, for example, \citet[]{Maio2007} shows that the cooling rate stays approximately the same between $10^{3}$ and $10^{4}$~K for gas with a metallicity below $Z=10^{-3}$, we justify the simplification below $10^{4}$~K by noting that cooling below $10^{4}$~K has a negligible effect on the formation and fragmentation of dense clouds as cooling in shocked gas and turbulent mixing layers is limited by numerical resolution rather than by radiative cooling \citep{Fujita2009, Gronke2018, 2020MNRAS.494L..27G}.
{We justify CIE assumption because past simulations show that the effects of non-equilibrium ionization (NEI) do not much boost high ion distributions even in shocked coronal gas \citep{KwakShelton2010, Armillotta2016, Cottle2018}.}
We do not include the effects of a metagalactic UV background radiation in our simulation, 
but we incorporate them when we post-process the simulations to compute the ion distributions (see Sect.~\ref{sec:metals}). The modification of the ionization fraction by a UV background would affect only the lower density gas that does not dominate the cooling.
\subsection{Starburst} \label{sec:sb}
In our study, we set up an instantaneous starburst of total stellar mass
$4\times10^{6}$~M$_{\odot}$ at the disk center. 
We use Stellar
Yields for Galactic Modeling Applications \citep[SYGMA][]{RItter2018} to
model the chemical ejecta and feedback from simple stellar
populations (SSPs). SYGMA is part of the open-source chemical evolution
NuGrid framework (NuPyCEE\footnote{http://www.nugridstars.org}). We
compute the average 
mechanical luminosities and the average metal ejection rates for
M$_{\rm SSP}=4\times10^6$~M$_{\odot}$. They are $L_{\rm
  SNII}=4\times10^{40}$~erg~s$^{-1}$ and $\dot{M}_{\rm
  SNII}=1.2\times10^{-3}$~M$_{\odot}$ yr$^{-1}$ for the initial
40~Myr, which is the lifetime of the smallest B star to go core
collapse Type II supernova (SNII), and $L_{\rm
  SNIa}=8\times10^{37}$~erg~s$^{-1}$ and 
  $\dot{M}_{\rm
  SNIa}=1.0\times10^{-5}$~M$_{\odot}$ yr$^{-1}$ at times $\ge40$~Myr
powered by Type Ia supernovae (SNIa). The metals produced by SNII and
SNIa are followed and advected separately.

To drive a constant-luminosity outflow, {during every time step
  $\Delta t$} we add mass ($\dot{M}_{\rm
  in}${$\Delta t$}) and
energy ($L_{\rm SNII}${$\Delta t$} and $L_{\rm SNIa}${$\Delta t$}) to a spherical source
region with a radius of 102.4~pc. We choose to increase the amount of
mass added from the SYGMA values to ensure that the temperature of hot gas in the
outflows is $3\times10^{7}$~K, which is far from the peak of the
cooling curve at $\sim10^{5}$~K{, but well below the value implied by
only accounting for the ejecta}. This additional mass accounts for
the mass evaporated off the swept-up shells in the absence of an
implementation of thermal heat conduction. Therefore, we use $\dot{M}_{\rm
  in} = 8.3\times10^{-2}$~M$_{\odot}$~yr$^{-1}$ for the SNII driven
outflow and $1.7\times10^{-4}$~M$_{\odot}$~yr$^{-1}$ for the SNIa
driven outflow. The total mass added for 1 Gyr
is only $3.48\times10^{6}$~M$_{\odot}$ which is less than 1$\%$ of
M$_{\rm disk}$. 
 \\
\subsection{Ion Analysis} \label{sec:ion}
We use the TRIDENT analysis tool \citep[]{Hummels2017} to calculate the ionization fractions of the species of interest based on the cell-by-cell density, temperature, and metallicity. First, the estimation for the number density of an element \textit{X} is 
\begin{equation}
    n_{X}=n_{H}~\frac{\mathrm{Z}}{~~\mathrm{Z_{\odot}}}~\left(\frac{n_{X}}{n_{H}}\right)_{\odot},
	\label{eq:nx}
\end{equation}
where Z is the metallicity from the simulation, and $(n_{X}/n_{H})_{\odot}$ is the solar abundance by number. 
Ionization fractions are pre-calculated over a grid of temperature,
density, and redshift in photoionization equilibrium (PIE) with the metagalactic UV background radiation by \citet[]{Haardt2012}, using the photoionization software CLOUDY \citep[]{Ferland2013}. Thus by linearly interpolating over the pre-calculated grid, TRIDENT returns the density of an ion, \textit{i}, of an element, \textit{X} as
\begin{equation}
    n_{X_{i}}=n_{X}f_{X_{I}},	
    \label{eq:nxi}
\end{equation}
where $f_{X_{I}}$ is the ionization fraction of the \textit{i}th ion. 

\begin{figure*}[!htp]\centering
\includegraphics[width=1\textwidth, scale=0.6]{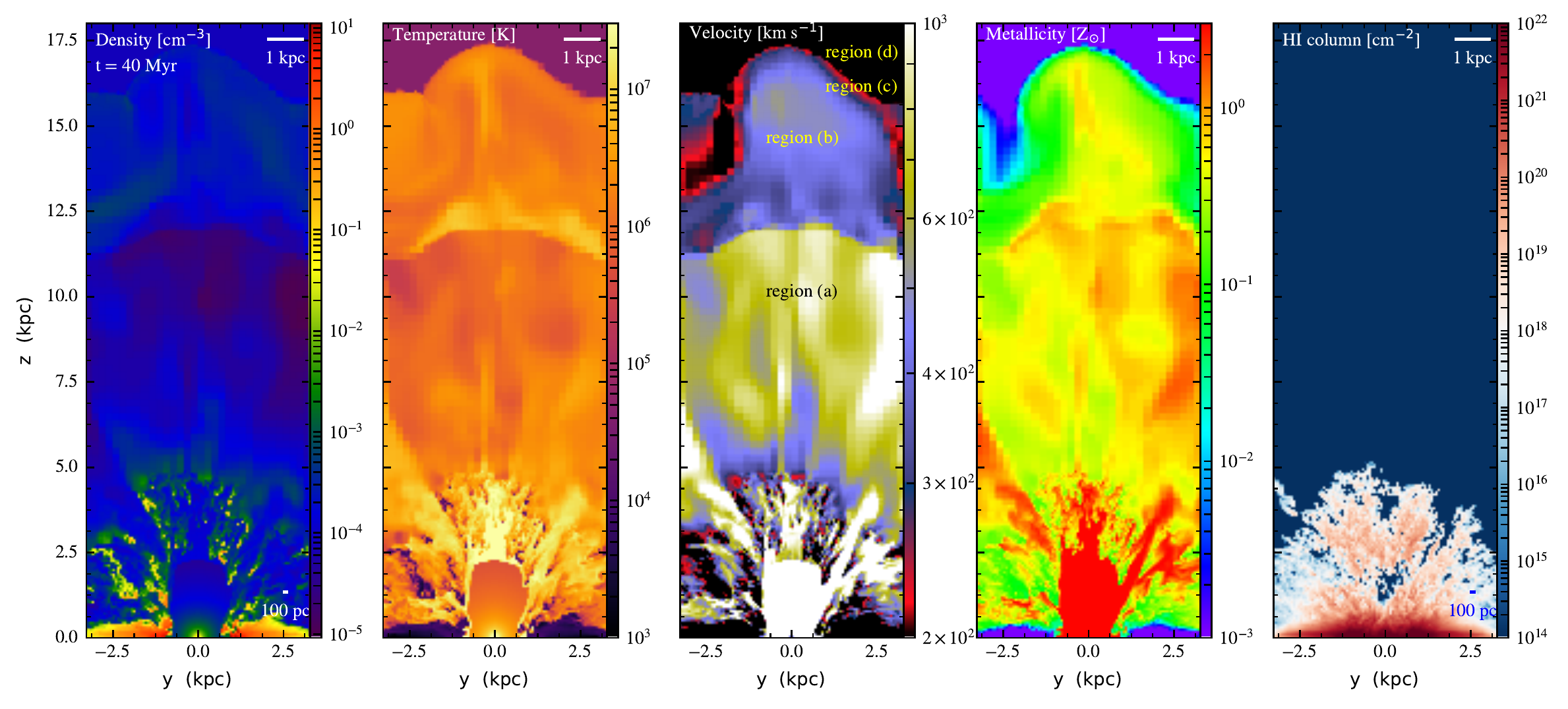}
\caption{Sliced density, temperature, velocity {magnitude}, and metallicity (\textit{from left to right}) distributions and  a projected hydrogen column density distribution along the $x$-axis (\textit{rightmost}) of the SNII-driven outflow at the box center ($y$-$z$ plane)  when the last SNII goes off at $t$=40~Myr. {The middle figure denotes region a) expanding SNII enriched gas, b) shocked SNII enriched gas, c) swept-up CGM, and d) the ambient CGM.} \label{fig:40}}
\end{figure*}

To generate an absorption profile along a ray through the simulation
box, the absorption produced by each grid cell is represented by a
single Voigt profile at its instantaneous velocity $v$, with a Doppler
$b$ parameter specified by the temperature in the cell. 
\section{Results} \label{sec:results}
Figure~\ref{fig:40}  shows density, temperature, total velocity, and
metallicity slices along the y-z plane at the disk center and a
neutral hydrogen (HI) column density distribution along $x$ axis in
the $y$-$z$ plane at t=40~Myr. The HI distribution is calculated with
Trident. 

The swept-up shell driven by repeated SNII explosions cools quickly
due to its high density. {Because it is expanding into a
  stratified atmosphere, it accelerates} and fragments into multiple
clumps and shells due to the Rayleigh-Taylor (RT)
instability. Figure~\ref{fig:40} shows that the hot, thermalized interior gas expands freely through
the fragments, forming a supersonic, energy-driven
outflow.  Kelvin-Helmholtz instabilities ablate the sides of these
fragments as the hot gas streams past them. This outflow continues to
shock the CGM, and a classic superbubble \citep{Weaver77} {forms
  in the CGM, as} seen in Figure~\ref{fig:40}: region (a) expanding
SNII enriched gas at v$\sim$400--1000 km s$^{-1}$, region (b) shocked
SNII enriched gas, region (c) swept-up CGM shell which is very thin and {light}
because there is not much to sweep due to its low density, 
and region (d) the ambient CGM beyond the outer shock front at
z$\sim$17~kpc.  Expanding SNII enriched gas and shocked SNII enriched
gas are divided at the inner shock front at z$\sim$12~kpc, and shocked
SNII enriched gas extends out to a contact discontinuity with the
CGM. 

In our simulations, the high-density, low-temperature fragments of
swept-up ISM material are not resolved after $t=40$~Myr with our refinement
   criteria of strong pressure gradients or the sound crossing time
   exceeding the cooling time.
They are Lyman Limit Systems (LLSs) and sub-Damped Lyman-alpha Absorbers
(DLAs) with $N_{HI}\gtrsim10^{18-20}$~cm$^{-2}$ that will likely
produce strong Mg{\sc ii} absorbers (see \textit{rightmost} figure in
Figure~\ref{fig:40}). The focus of this study is instead on weak Mg{\sc ii}
absorbers that are observed to be associated with sub-LLSs with
$N_{HI}\lesssim10^{17}$~cm$^{-2}$. These unresolved swept-up ISM
fragments in the outflow quickly mix with the surrounding hot, metal
enriched gas, but the total amount of disk gas mixed in the outflow is
only 3--5$\%$ of the disk mass initially placed on the grid. We also
note that the powerful SNII driven outflow leaves the box starting at
$t\sim20$~Myr;  by $t=40$ to 300~Myr, 38\%
to  58\% of the metal-carrying gas has left the box. 
 
 After the last SNII goes off at $t=40$~Myr, SNIa's drive the outflow,
 but with a mechanical luminosity that is more than two orders of
 magnitude smaller. SNIa-enriched gas expands at $v\sim400$~km
 s$^{-1}$ through the tunnel created by the previous SNII outflow, but
 by $t\sim80$~Myr the disk gas being pushed aside by the SNII outflow
 flows back to the central source region, blocking the passage for
 SNIa-enriched gas. Meanwhile, the shocked SNII-enriched gas (region
 b) near the inner shock front ($z\sim12$~kpc) begins to descend
 toward the disk, while the outer shock front (the outer edge of region c) keeps moving at
 $v\sim400$~km~s$^{-1}$ in the CGM and soon leaves the box. By
 $t\sim100$~Myr,  descending shocked SNII enriched gas accumulates at
 the inner shock front and cools to form denser, cool shells that eventually fragment by RT instability. 
 
\begin{figure*}[!htp]\centering
\includegraphics[width=1\textwidth, scale=0.6]{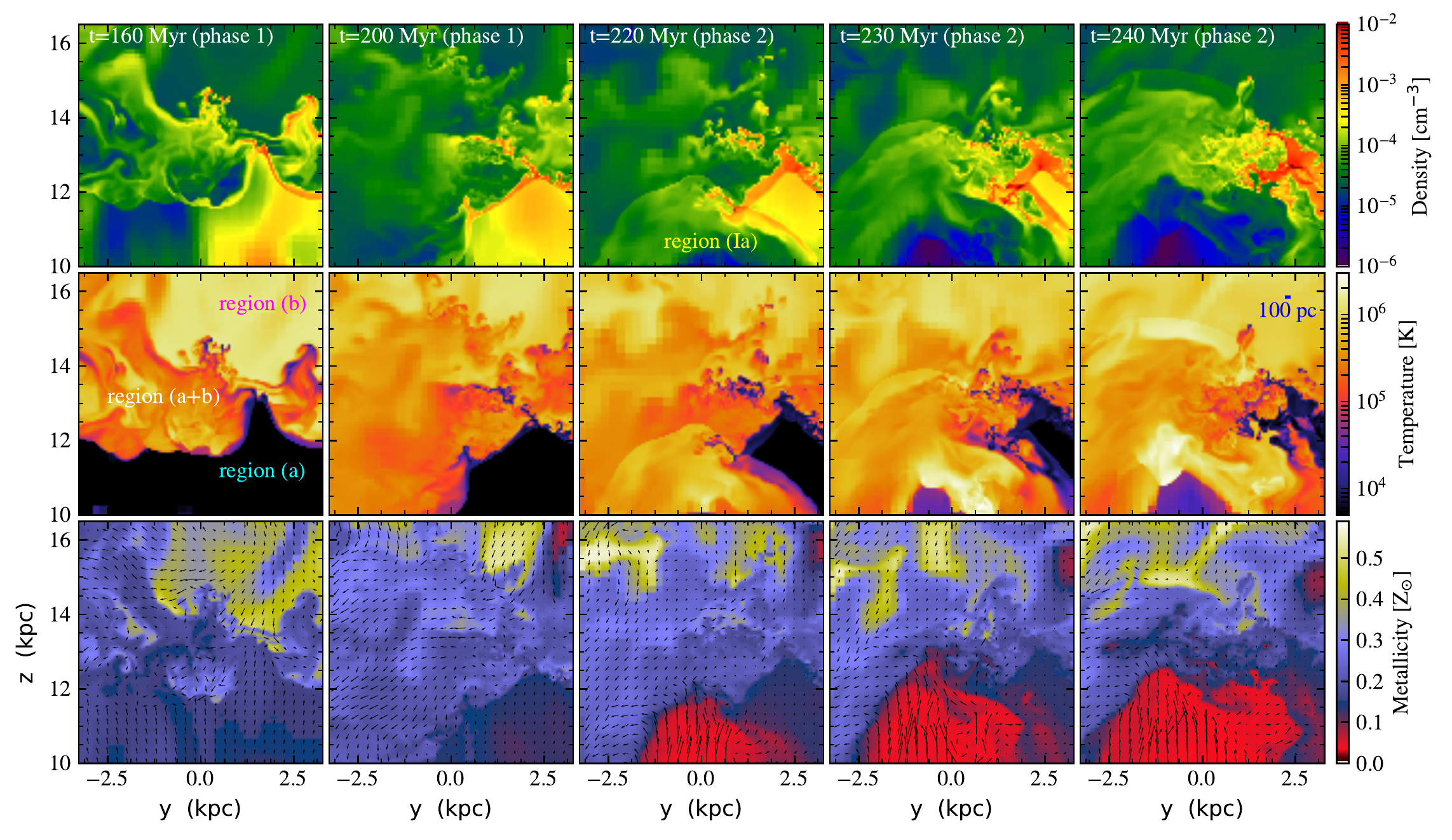}
\caption{Sliced density (\textit{top}), temperature
  (\textit{middle}), and metallicity  (\textit{bottom}) distributions
  of cool, dense clouds at $x=+1.42$ kpc from the disk center in
  $y$-$z$ plane at \textit{phase 1} (t=160 and 200 Myr) and
  \textit{phase 2} ($t=220$, 230, and 240 Myr) \textit{from left to
    right}.  \textit{Phase 1} formation begins when descending shocked
  SNII enriched gas ({\textit{region b}}) collides with the expanding SNII enriched gas ({\textit{region a}}) at
  the inner shock front, and \textit{phase 2}  formation begins when
  SNIa driven outflow ({\textit{region Ia}}) rams into the rest of the SNII enriched gas and
  the clouds made at \textit{phase 1}.  The arrows in \textit{bottom} figures show the direction of gas flow with v$_{max}=429$~km~s$^{-1}$. \label{fig:cloud}}
\end{figure*}

\begin{figure}[!hbp]\centering
\includegraphics[width=1.1\columnwidth]{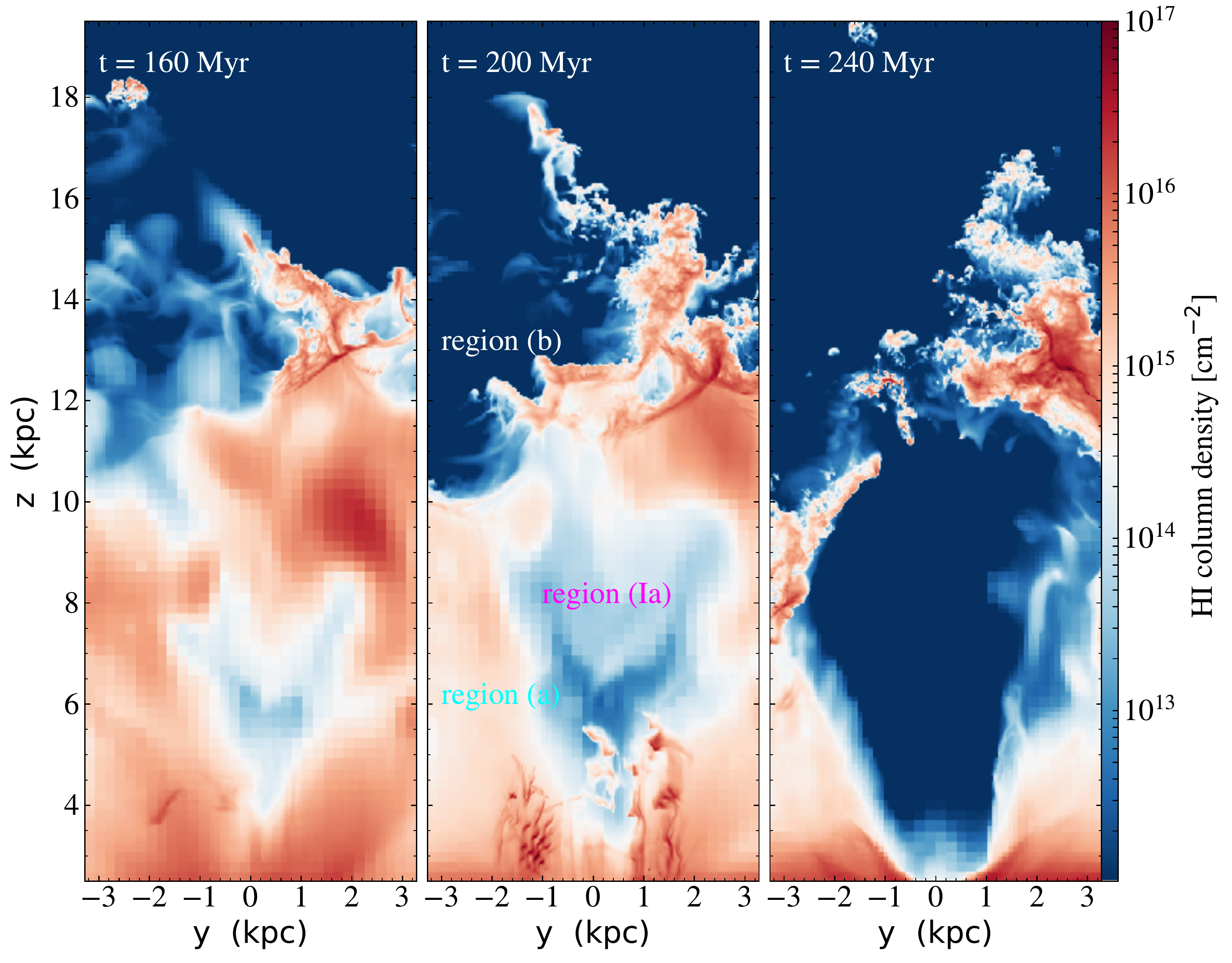}
\caption{Projected neutral hydrogen distributions at $t=160$
  (\textit{left}), 200 (\textit{middle}), and 240~Myr
  (\textit{right}), along the $x$-axis in the $y$-$z$ plane. {SNIa driven outflow is visible as a cavity  (\textit{region Ia})} \label{fig:HI}}
\end{figure}

 The sliced density distribution in the $y$-$z$ plane at $x=+1.42$~kpc
 from the disk center at $t=160$~Myr (\textit{left} in
 Figure~\ref{fig:cloud}) shows the formation of such fragments in the
 form of clumps and filaments. They are also visible as clumps and
 filaments in a projected distribution of neutral hydrogen along the
 $x$-axis at $t=160$ Myr (\textit{left} in Figure~\ref{fig:HI}). These
 clumps and filaments will potentially produce weak Mg{\sc ii} absorbers (we
 discuss our ion analysis in the next section). We call this process
 \textit{phase 1} formation.  They are made of SNII enriched
 outflow gas and their metallicity is $\sim0.1$--$0.2 Z_{\odot}$. The
 size of clumps and the thickness of filaments are $\sim$100~pc.
    This size may be limited by our numerical 
    resolution of 12.8~pc
\citep{Fujita2009,Gronke2018}. We discuss the effects of resolution
further in Sec.~\ref{sec:resolution}. 

\begin{figure*}[!htp] \centering
\includegraphics[width=0.62\textwidth , scale=0.5]{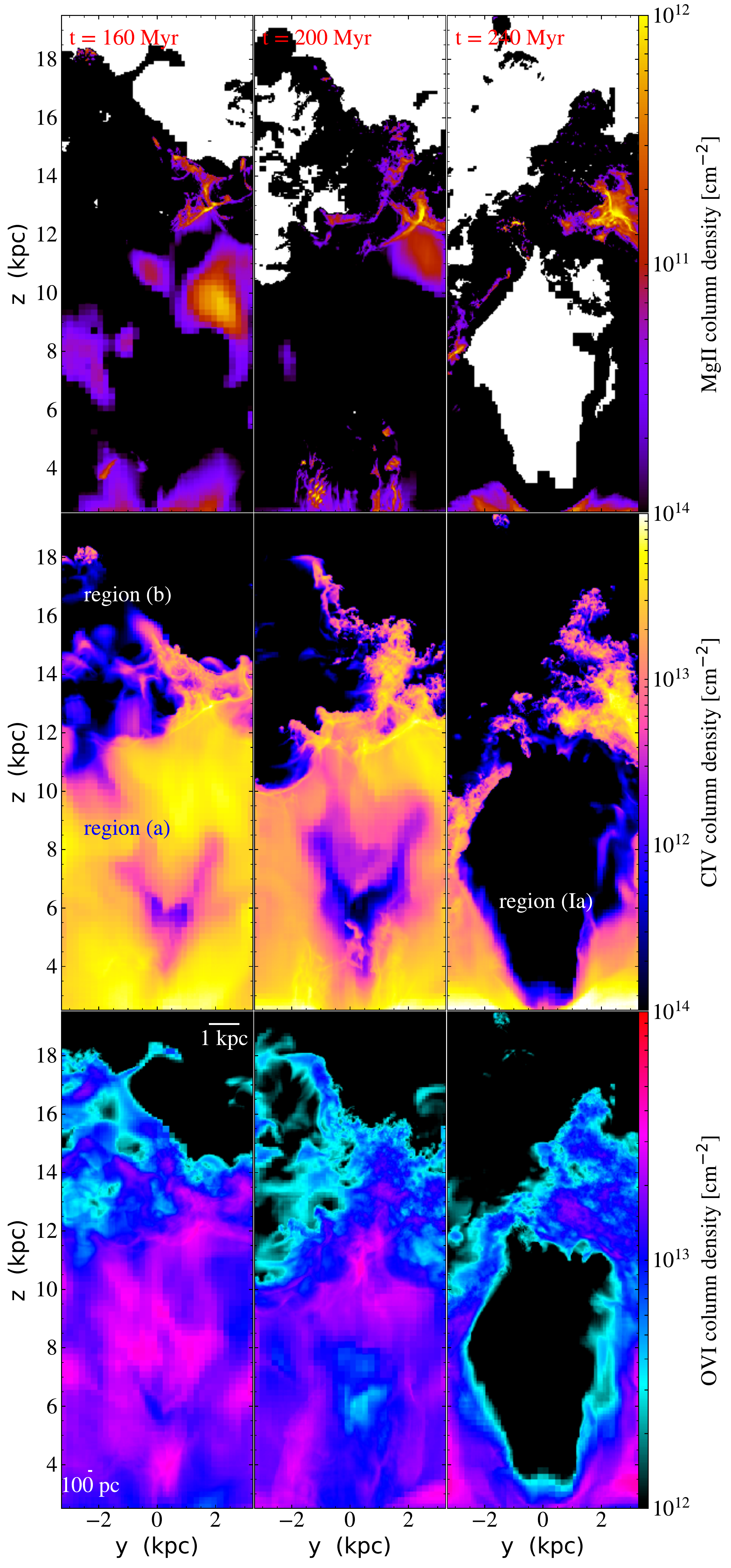} 
\caption{Projected Mg{\sc ii}  (\textit{top}), C{\sc iv}  (\textit{middle}), and
  O{\sc vi} density (\textit{bottom}) distributions at $t=160$
  (\textit{left}), 200 (\textit{middle}), and 240~Myr
  (\textit{right}), along the $x$-axis in the $y$-$z$ plane.  \label{fig:CIVOVI}}
\end{figure*}

Shortly after $t=160$~Myr, a superbubble created by repeated SNIa
explosions blows out of the dense ISM and SNIa enriched gas regains a
tunnel for expansion, forming a SNIa driven outflow traveling at $v\sim400$--500~km s$^{-1}$.  In the projected distribution of neutral hydrogen at $t=200$~Myr (\textit{middle} panel in Fig.~\ref{fig:HI}),  fragments of swept-up ISM after blowout are visible framing a tunnel for outflow, and hot, low-density SNIa enriched gas in the outflow is seen as a cavity with $N_{HI}\lesssim10^{13}$~cm$^{-2}$ {(we define SNIa enriched gas as region Ia)}. 

By $t=220$~Myr, this SNIa driven outflow (region Ia) expands into the cooled SNII
enriched gas and the clumps and filaments of shocked SNII
enriched gas (region b), shocking and sweeping
them and forming more clumps and filaments. Figure~\ref{fig:cloud}
shows such a process clearly in a selected region at $z>10$~kpc. These
are potential candidates for weak Mg{\sc ii} absorbers, too: we call this
process \textit{phase 2} formation. Their metallicity and size
are likewise $\sim0.1$--$0.2 Z_{\odot}$ and
$\sim$100~pc. Hotter and lower-density shocked SNII enriched gas
carries more metals ($Z\sim0.4$--$1 Z_{\odot}$) and lies above
$z\sim14$ kpc. 

The SNIa driven outflow continues to shock and sweep
gas as well as clumps and filaments to the sides, and by
$t\sim300$~Myr, all the clumps and filaments as well as 58\% of SNII
outflow gas and 8\% of SNIa outflow gas have left the box. Then, there
is only very low-density gas with n$_{H}<10^{-4}$ cm$^{-3}$ left above
the disk in the box. The metallicity of SNIa enriched gas is $Z \ll
0.1 Z_{\odot}$ as the metal production rate is about two orders of magnitude smaller than that of SNII, so it is still too early for any significant enrichment by SNIa. We stopped computing at  $t\sim450$~Myr. 
\begin{figure*}[htp!]\centering
\includegraphics[width=1.05\textwidth,scale=0.6]{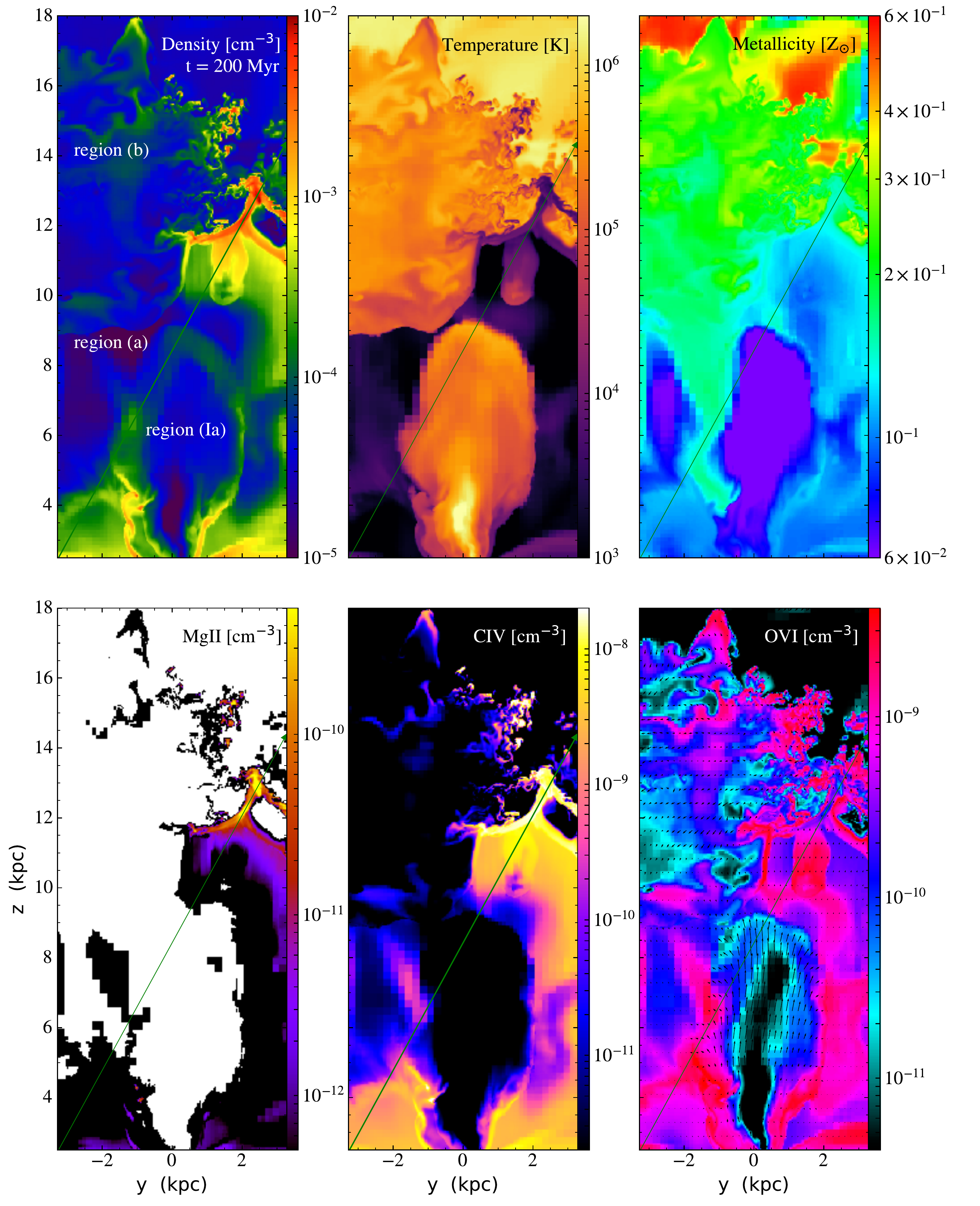}
\caption{Sliced density, temperature, metallicity (\textit{top from left to right}), and Mg{\sc ii}, C{\sc iv}, and O{\sc vi} (\textit{bottom from left to right}) density distributions at x=+1.92~kpc from the disk center in the y-z plane, at t=200~Myr. A line of sight from [x,y,z]=[+1.92~kpc, -3.28~kpc, +2.45~kpc] to [+1.92~kpc, +3.28~kpc, +14.4~kpc] is shown by a \textit{green} line \label{fig:t2000spec}. The arrows in the \textit{bottom right} figure show the direction of the gas flow with v$_{max}$=353~km~s$^{-1}$.}
\end{figure*}

With a realistic star formation history with multiple star clusters scattered in time and place, we expect \textit{phase 1} and \textit{phase 2} formation to be repeated in time and place to produce more clumps and filaments. 
We will test this scenario in a larger simulation box in our next paper. 

\section{Weak Mg{\sc ii} Absorbers and C{\sc iv}/O{\sc vi} absorbers}\label{sec:metals}
We calculate the fractions of H{\sc i}, H{\sc ii}, Mg{\sc ii}, C{\sc iv}, and O{\sc vi} ions using Trident as described in Section~\ref{sec:ion} by assuming all the gas in our simulation to be in photoionization equilibrium (PIE) with the UVB radiation at a given redshift \citep{Haardt2012}. 
Our simulation does not include the effects of UVB radiation, so for example, expanding SNII enriched gas {tends to overcool to lower temperature, $\lesssim10^{4}$ K. However, this overcooled, low-density ($\leq10^{-4}$ cm$^{-3}$) gas contributes very little to the total ion budgets, and denser clouds that produce Mg{\sc ii} absorbers are self-shielded to the surrounding UVB radiation as long as n$_{H}\sim5\times10^{-3}$ cm$^{-3}$ at z=2 \citep{Rahmati2013}. Thus overcooling will not significantly affect our analysis (see Appendix).  }
\subsection{Overview}
Figure~\ref{fig:CIVOVI} shows projected density distributions of Mg{\sc ii}, C{\sc vi}, and O{\sc vi} ions along the x-axis in the y-z plane at 
t=160, 200, and 240 Myr, and Figure~\ref{fig:t2000spec} shows sliced density, temperature, metallicity, Mg{\sc ii}, C{\sc vi}, and O{\sc vi} ion density distributions at x=+1.92~kpc from the disk center in the y-z plane at t=200~Myr.  
{This sight line was selected as an example with a large pathlength through low ionization gas.}

The clumps and filaments have hydrogen number densities, n$_{H}=10^{-3}$ to $10^{-2}$ cm$^{-3}$, and their sizes/thickness, $\sim$100 pc, which is the smallest scale our simulation can resolve, as discussed in Section ~\ref{sec:results}. 
Individual weak Mg{\sc ii} absorbers seem to survive for $\sim60$ Myr, {before they are mixed and diluted with the surrounding, warmer, lower-density gas}, but they are continuously produced through \textit{phase 1} to \textit{phase 2} formation 
for over 150 Myr from a single instantaneous starburst source. Weak Mg{\sc ii} absorption with $N_{MgII}>10^{11}$ cm$^{-2}$ is also found in a blob of gas that carries a swept-up ISM shell fragment in the expanding SNII enriched gas seen 
at e.g. [y,z]=[+2 kpc, 10 kpc] (see \textit{top left} figure in Figure~\ref{fig:CIVOVI}) and in fragmented shells of ISM swept-up by the SNIa driven outflow at e.g. z=2--4 kpc (see \textit{top middle} figure in Figure~\ref{fig:CIVOVI}). The blob has cooled slowly without fragmentation, and its size is $\gtrsim$ kpc. It is expanding into the \textit{phase 1} shells in region (b) above, but SNIa driven outflow will shock and sweep up {expanding SNII enriched gas including the blob in region (a) and the \textit{phase 1} shells in region (b) } to produce \textit{phase 2} shells and fragments (see \textit{top right} figure in Figure~\ref{fig:CIVOVI}).  

Higher ion absorbers are found in region (a) where expanding SNII enriched gas cools and in region (b) where shocked SNII enriched gas cools in \textit{phase 1} and \textit{phase 2}. 
In both cases, the hydrogen number density of the absorbers is n$_{H}\sim$a few $\times10^{-4}$ cm$^{-3}$, but the absorbers in region (a) extend over 1-4 kpc while the absorbers in region (b) are smaller, 500 pc--1 kpc. The sizes of high ion absorbers agree with the observed estimates for C{\sc iv} absorbers by \citet[]{Misawa2008} and \citet[]{Shaye2007}. They are $\sim$100 pc - 5 kpc in a sub-LLS ($10^{14.5}<$$N_{HI}<10^{16}$ cm$^{-2}$)
or Ly$\alpha$ forest environment ($N_{HI}<10^{14.5}$ cm$^{-2}$).
\begin{figure*}[!h]\centering
\includegraphics[width=0.8\textwidth, scale=0.8]{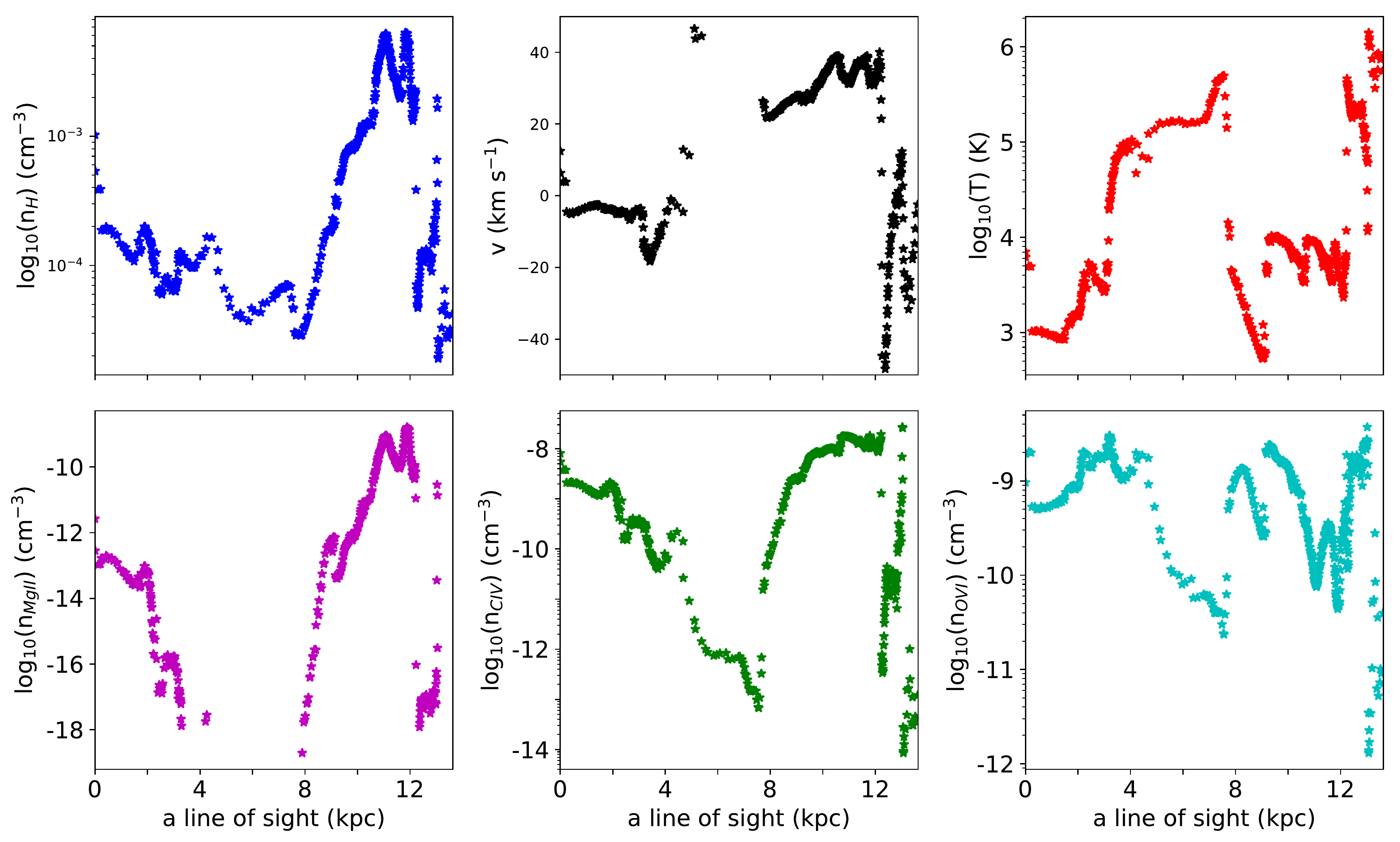}
\caption{Hydrogen density (\textit{top left}), sightline velocity (\textit{top middle}), temperature (\textit{rop right}), Mg{\sc ii} density (\textit{bottom left}), C{\sc iv} density (\textit{bottom middle}), and O{\sc vi} density (\textit{bottom right}) distributions along the line of sight from [x,y,z]=[+1.92~kpc, 0~kpc, 2.45~kpc] to [+1.92~kpc, 6.55~kpc, 14.4~kpc] (\textit{green line} in Figure~\ref{fig:t2000spec}) at t=200~Myr.  \label{fig:jf2000-q}}
\end{figure*}
\begin{figure*}[!h]\centering
\includegraphics[width=0.8\textwidth, scale=0.8]{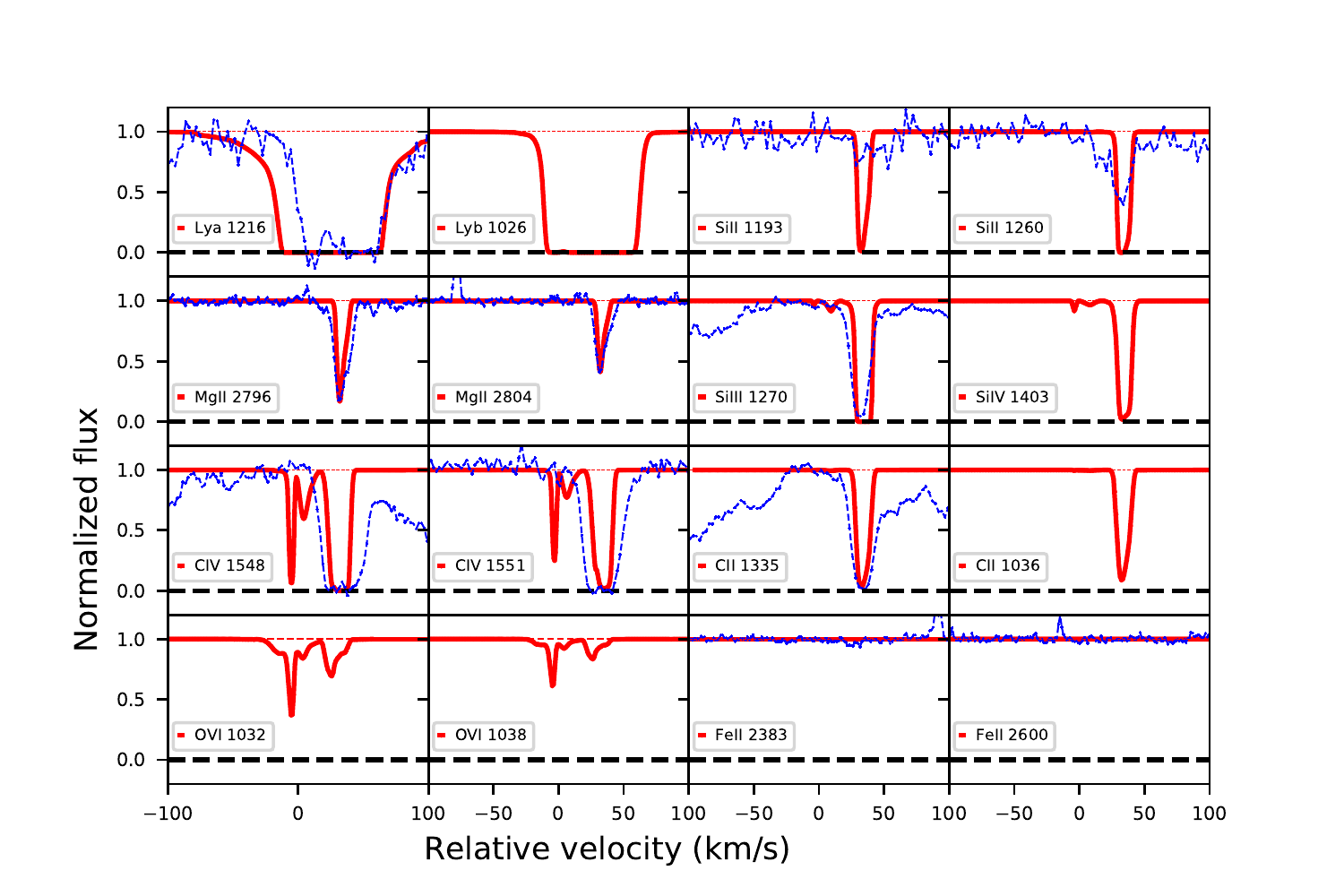}
\caption{Mock spectra along the line of sight (\textit{green line} in Figure~\ref{fig:t2000spec}) at t=200~Myr, compared to the observed profiles of system 3 at z=1.75570 \citep[\textit{blue dashed line},][]{Misawa2008}. \label{fig:spec2000-1}}
\end{figure*}
\begin{figure*}[!ht]
{\centering
\includegraphics[width=0.99\textwidth, scale=0.6]{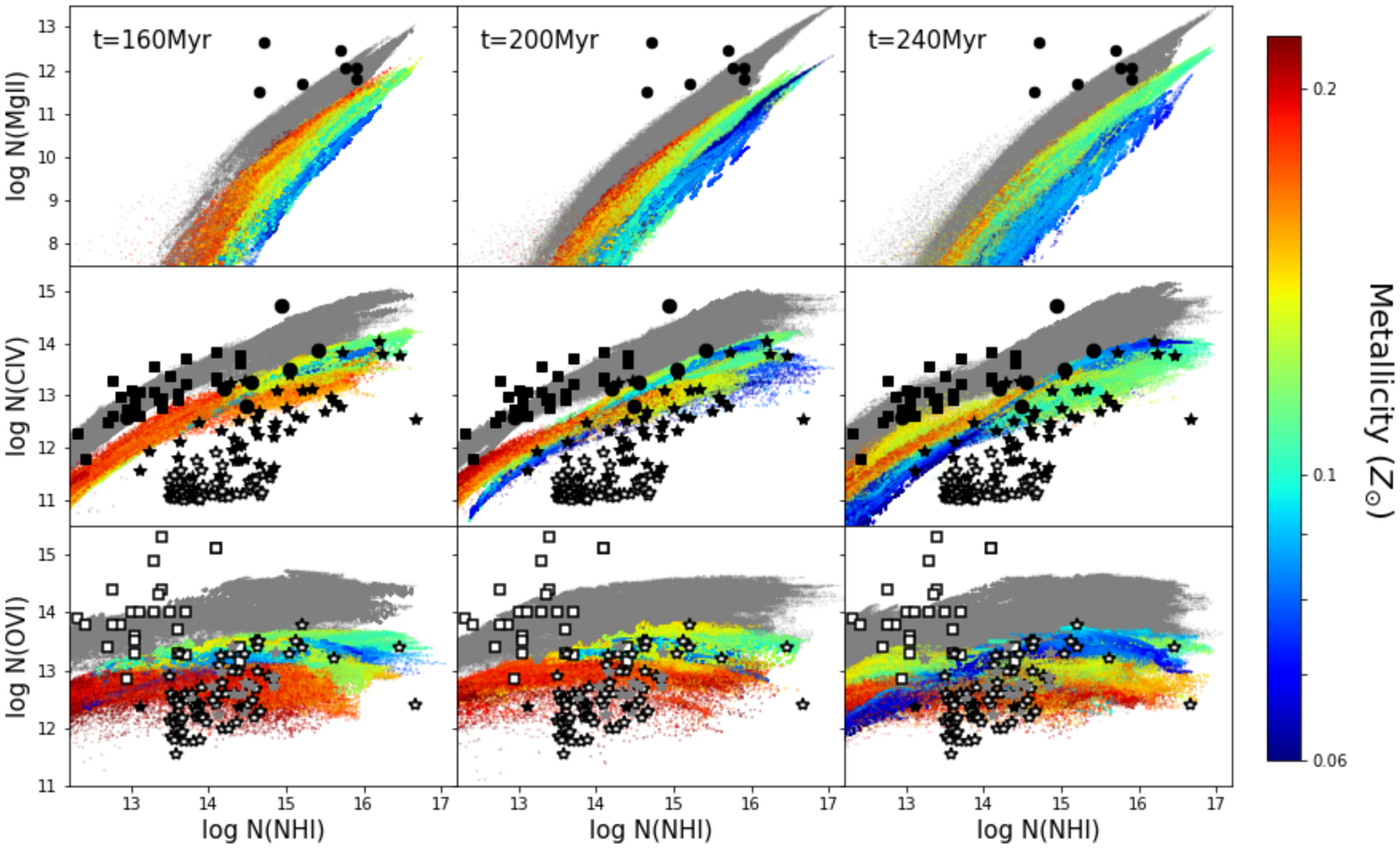}
\figcaption{Mg{\sc ii} (\textit{top row}), C{\sc iv} (\textit{middle row}), and O{\sc vi} (\textit{bottom row}) versus H{\sc i} column densities \added{in sightlines parallel to each of the three cardinal axes} at t=160 (\textit{left column}), 200 (\textit{middle column}), and 240 Myr (\textit{right column}) with different colors indicating Mg{\sc ii}, C{\sc iv}, and O{\sc vi} density-weighted metallicities, to be compared to the observed Mg{\sc ii}/C{\sc iv} clouds by \citealt{Misawa2008}~(\textit{circle}) and the observed C{\sc iv}/O{\sc vi} observations by \citealt{Shaye2007}~(\textit{square}) and
{\citealt{DOdorico2016}~(\textit{star, but gray star for detection of only one member of the doublet}) . Note O{\sc vi} densities from \citealt{Shaye2007}~(\textit{open square}) and C{\sc iv} and O{\sc vi} densities from \citealt{DOdorico2016}~(\textit{open star}) are upper limits (\textit{open square}). Grey points indicate ion versus HI column density distributions expected when all the gas in our simulation is assumed to have solar metallicity.
} 
\label{fig:scatter}} 
}
\end{figure*}
C{\sc iv} absorbers in region (b) are clumpy and filamentary and some surround weak Mg{\sc ii} absorbers, so both of them arise from the same clumps and filaments created in \textit{phase 1} and \textit{phase 2} formation. However, C{\sc iv} ions in these clumps and filaments survive longer than Mg{\sc ii} ions by another 20-30 Myr. Our simulations suggest clouds that produce Mg{\sc ii} absorbers also produce C{\sc iv} absorbers, and Mg{\sc ii} absorbers probe the densest parts of the clouds while C{\sc iv} absorbers extend out to more diffuse, larger regions. In the process of mixing, the regions that produce Mg{\sc ii} absorption disappear first due to dilution, so {our simulations agree with a picture proposed by \citet{Shaye2007} that expanding Mg{\sc ii} absorbers with high metallicity (Z$\lesssim$Z$_{\odot}$) produce C{\sc iv} absorbers.}

O{\sc vi} absorbers in region (b) appear even more diffuse, arising in larger ($\gtrsim1$ kpc) clouds, that are not necessarily the same clouds that host Mg{\sc ii} and C{\sc iv} clouds nor in the same phase. On the other hand, most O{\sc vi} absorbers in region (a) arise from the same clouds that host C{\sc iv} absorbers in the same phase. 

We find that 1-3$\%$ of high ions {by mass are from collisional ionization, and they are found in coronal O{\sc vi} absorbers in region (b).
This is consistent with observational analyses that photoionization dominates in sub-LLS and Ly$\alpha$ forest environments at intermediate to high redshift.} \citep[e.g.][]{Simcoe2004, Shaye2007, Lerner2016}. 

Figure~\ref{fig:jf2000-q} shows physical values along a line of sight through the simulation box, which is noted in \textit{green} in Figure~\ref{fig:t2000spec}, and Figure~\ref{fig:spec2000-1} shows mock spectra created along the sightline with Trident. Along the sightline, there are two Mg{\sc ii} absorbers which correspond to two peaks in Figure~\ref{fig:t2000spec} and in the \textit{bottom left} plot of Figure~\ref{fig:jf2000-q}. They are shocked cooling shells in region (b) and {are only separated by a small velocity in the spectrum, despite their spatial separation} ($\Delta$v$\sim$2~km s$^{-1}$ at v$\sim$38~km s$^{-1}$), which is visible in the absorption profile {as a slight asymmetry}(Figure~\ref{fig:spec2000-1}).  

The same shells produce C{\sc iv} absorption, but no O{\sc vi} absopbtion. O{\sc vi} absorbers in region (b) are in a different, coronal phase. C{\sc iv} absorbers in region (a) are $>$ few kpc in size: one is at z$\sim$2.5 kpc with a positive velocity (v$\sim$10~km s$^{-1}$), one is at z$\sim$2.5-4 kpc with a nagative velocity (v$\sim$-5~km s$^{-1}$), one is at z$\sim$10--11 kpc (cooler, v$\sim$30~km s$^{-1}$) and the other is at z$\sim$11--12 kpc (warmer, v$\sim$40~km s$^{-1}$), both below the cooling shell (v$\sim$38~km s$^{-1}$). The first two absorbers produce the double absorption profiles in Figure~\ref{fig:spec2000-1}, and the last two absorbers produce the saturated absorption profile at v=20--45~km~s$^{-1}$, together with C{\sc iv} absorbers in region (b).

O{\sc vi} absorbers in region (a) arise from the same cold clouds, producing two sets of double absorption profiles, but the sightline is also going through a turbulent mixing layerv{of} swept-up shells by the SNIa driven outflow at z=5--7 kpc. Its temperature is $\gtrsim$10$^{5}$~K. The signal is buried in the double absorption profiles at v$\sim$10~km s$^{-1}$. The O{\sc vi} absorber in region (b) is coronal and turbulent with v$\sim$-10--40~km s$^{-1}$, but is weak compared with the other O{\sc vi} absorbers. 

{We note that some SNII outflow gas in region a cools to temperature below $10^{4}$ K by t$\gtrsim$200 Myr, however, this overcooled, low-density ($\leq10^{-4}$ cm$^{-3}$) gas only makes a little contribution to C{\sc iv} and O{\sc vi} column column densities (see Appendix). }

Figure~\ref{fig:spec2000-1} also shows that our mock spectra reproduce qualitative features of the observed profiles of {observed} weak Mg{\sc ii} system 3 at z=1.75570 {published in} \citet{Misawa2008}.

\subsection{Comparison to observations}
\subsubsection{Column densities and metallicities}
In Figure~\ref{fig:scatter}, we show the relations between ion column densities and HI column densities in our simulation {at t=160, 200, and 240 Myr in sightlines parallel to each of the three cardinal axes}, and compare them with the observed relations. The colors indicate {Mg{\sc ii}, C{\sc iv}, and O{\sc vi} density weighted metallicities respectively}. {Effective lower limits to the Mg{\sc ii}, C{\sc iv}, and O{\sc vi} column densities are $3.5\times10^{8}$, $7.5\times10^{8}$, and $4.7\times10^{9}$ cm$^{-2}$ in our simulations.} Mg{\sc ii}, C{\sc iv}, and O{\sc vi} absorbers in our simulation are enriched to Z=0.1-0.2 Z$_{\odot}$  by SNII from an instantaneous starburst, as the SNIa contribution is negligible at this point. 

\textit{Top} figures in Figure~\ref{fig:scatter} show that sightlines with higher metallicities have higher Mg{\sc ii} column densities at given HI column densities, and they are compared to 
the Mg{\sc ii}-HI observations from three Mg{\sc ii} absorbers at 
z$\sim1.7$ from \citet{Misawa2008} and four Mg{\sc ii} absorbers at lower redshift (z=0.65-0.91) from \citet{Charlton2003} and \citet{Ding2005}.  
The Mg{\sc ii} column densities in our simulation are up to 1 order of magnitude smaller than the observed values at the given HI column densities, $N_{HI}>10^{15}$ cm$^{-2}$ (i.e. sub-LLS).  This is mainly because our simulation can only be run up to $\sim300$ Myr before most gas leaves the grid: metal mixing and cooling should be computed for a much longer duration, also including the SNIa metal contribution. {Moreover, we set the initial metallicity of our dwarf disk and halo gas to be Z=$10^{-3}$ Z$_{\odot}$, to study the effects of metal contribution by our simulated starburst alone. Thus, we are likely underestimating the metallicities of Mg{\sc ii} absorbers.}
If we assume that all the gas in our simulation box has a solar metallicity, the boosted Mg{\sc ii} column densities (\textit{grey} points in Figure~\ref{fig:scatter}) agree more with the observed values. 

At lower $N_{HI}<10^{15}$ cm$^{-2}$(i.e. sub-LLS to Ly$\alpha$ forest), there is no dense cloud formation in our simulation thus no Mg{\sc ii} clouds with $N_{MgII}>10^{11}$ cm$^{-2}$. There are two Mg{\sc ii} absorbers with $N_{HI}<10^{14.5}$ cm$^{-2}$ at z$\sim$2 \citep{Misawa2008} and their Mg{\sc ii} column densities {are larger than predicted by our simulations for sightlines with this $N_{HI}$} by two orders of magnitude. This might also be due to {lower} metal enrichment in our simulation. The estimated metallicities for the two absorbers are very high, Z=0.63-0.79 and even super solar, Z$>7.9$ Z$_{\odot}$  respectively. {This discrepancy could also be related to the limited simulation resolution.} We hope to study the possible formation of super solar, weak Mg{\sc ii} clouds with our future global simulations. 

Simulated C{\sc iv} column density distributions appear to agree better with the observed column densities of C{\sc iv} absorbers that are found in the same sightlines with the Mg{\sc ii}  absorbers studied by \citet{Misawa2008}. These C{\sc iv} absorbers are in sub-LLS environments, and have similar metallicities, Z=0.1-0.3 Z$_{\odot}$ to our simulation values, except for one absorber with Z=0.8 Z$_{\odot}$: this metal rich C{\sc iv} absorber is in a structure related to the super solar, weak Mg{\sc ii} absorber with Z$>7.9$ Z$_{\odot}$. 

On the other hand, our simulated C{\sc iv} column densities {are smaller than those} of the C{\sc iv} absorbers studied by \citet{Shaye2007}: the disagreement is by an order of magnitude. {This is probably because these absorbers are selected for the high metallicities, Z$\sim$ Z$_{\odot}$.} They are found in Ly$\alpha$ forest environments and are smaller in size ($\sim$500 pc -- 1 kpc). In our simulation, smaller C{\sc iv} clouds {are found in region (b) and arise from the same clouds that currently host or used to host even smaller, weak Mg{\sc ii} absorbers} in sub-LLS to Ly$\alpha$ forest environments. Our metallicity boosted values better agree with the observations (Figure~\ref{fig:scatter}). The upper limits for O{\sc vi} column densities associated with the observed C{\sc iv} absorbers \citep{Shaye2007} are also above what our simulation predicts, and lie in the metallicity boosted grey area, just like most of the observed weak Mg{\sc ii} and C{\sc iv} absorbers.  There is no {other} information about their physical properties available. 

{The observed C{\sc iv} column densities by \citet{DOdorico2016} appear to agree with our simulated values at $N_{HI}>10^{14.5}$ cm$^{-2}$, however, they are much lower than our simulated values, by up to one order of magnitude, at $N_{HI}<10^{14.5}$ cm$^{-2}$. 
These C{\sc iv} absorbers are observed at a higher redshift, z$\sim2.8$, and the majority of them have their estimated metallicities between $10^{-2.5}$ Z$_{\odot}$ and $10^{-2}$ Z$_{\odot}$, much lower than our simulated values.
There is no information about the physical properties available for the C{\sc iv} and O{\sc vi} absorbers observed by \citet{DOdorico2016}. The data for O{\sc vi} column densities are mostly upper limits except three detections of which one shows a very weak C{\sc iv} line and another shows none. Out of 15 O{\sc vi} possible detections with single lines, six of them do not show an associated C{\sc iv} line. Despite the estimated low metal contents, the observed O{\sc vi} column densities and their upper limits appear to agree better with our simulated values at all HI environments. The sizes and thermal properties are unknown for these C{\sc iv} and O{\sc vi} absorbers.


We note that the observed estimates and upper limits for C{\sc iv} and O{\sc vi} column densities at given HI column densities vary over 4 orders of magnitudes. {This may be due to the presence of HI dominated gas in observed sightlines which originate in regions that are not covered by our simulations.} However, for Mg{\sc ii} absorbes and associated C{\sc iv} absorbers, a major reason for the discrepancy seems to be a lack of metal enrichment {as well as the low initial metallicity of disk and halo gas} in our simulation.} We speculate that galactic outflows from repeated bursts of star formation for a longer duration ($\sim1$ Gyr) will eventually create high-metallicity, complex structures of multiphase gas. 

\begin{figure*}[!ht]\centering
\includegraphics[width=0.7\textwidth, scale=0.6]{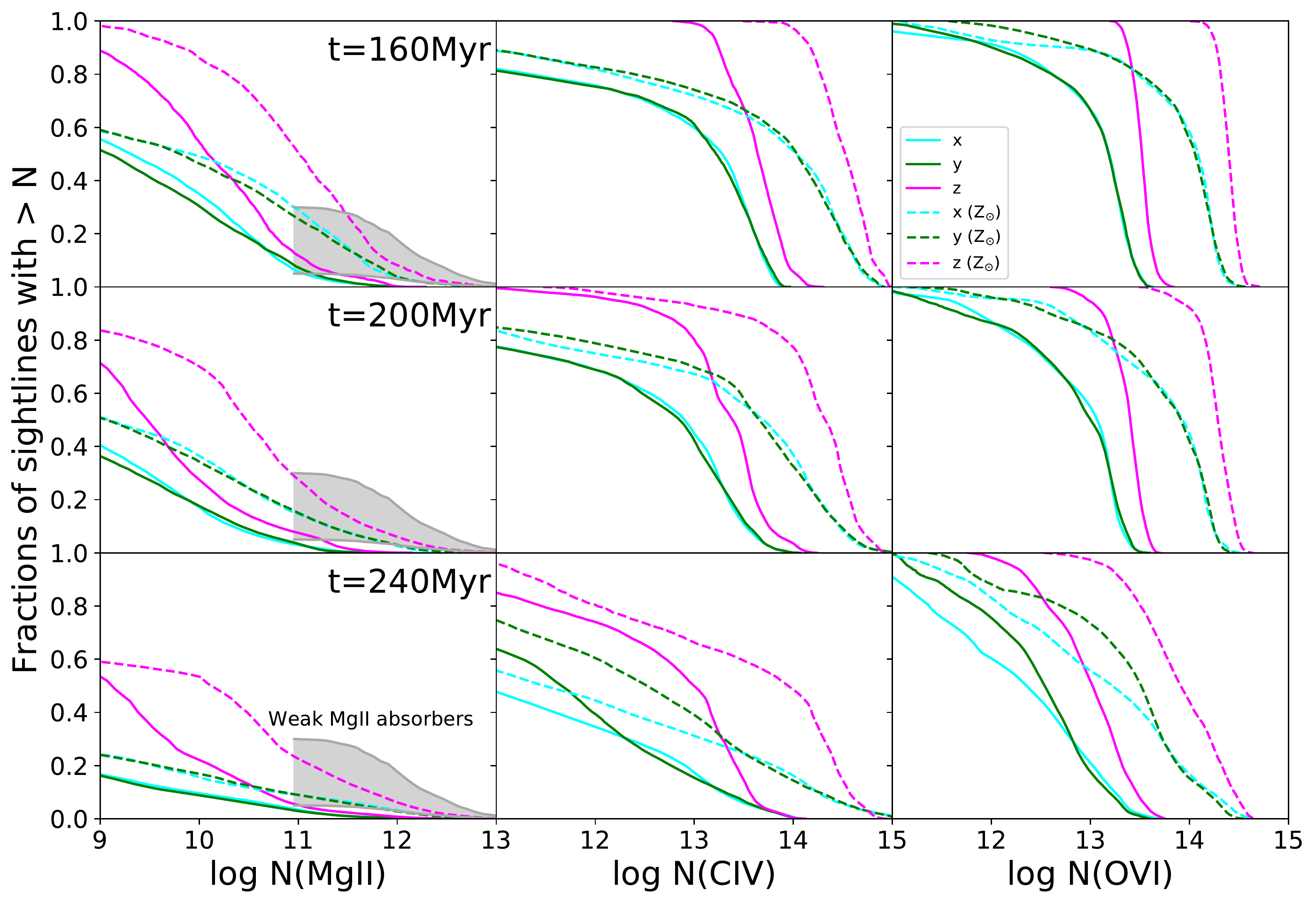}
\caption{The Mg{\sc ii} (\textit{left}), C{\sc iv} (\textit{middle}), and O{\sc vi} (\textit{right}) covering fractions as functions of column densities along each of the three cardinal axes at t=160 (\textit{top}), 200 (\textit{middle}) and 240 Myr (\textit{right}). All sightlines between z=2.5~kpc (the disk edge) and 17.5 kpc (virial radius) are included.  The \textit{dasremote.net.ed.ac.ukhed lines} show the covering fractions when all the gas is assumed to have solar metallicities. The \textit{grey region} indicates estimated fractions of sight lines as a function of Mg{\sc ii} column densities when we assume that the observed weak Mg{\sc ii} clouds at various redshifts \citep{Rigby2002, Charlton2003, Ding2005, Misawa2008, Narayanan2008} cover 5--30$\%$ of a halo. The observed Mg{\sc ii} column densities are $\geq10^{11}$ cm$^{-2}$.\label{fig:coverf}}
\end{figure*}
\subsubsection{Covering fractions}
Figure~\ref{fig:coverf} shows fractions of sight lines that occupy our simulation box above the galactic disk and within the virial radius as functions of Mg{\sc ii}, C{\sc iv}, and O{\sc vi} column densities along x, y, an z axes at three different times. 
%
The \textit{Grey} region in Figure~\ref{fig:coverf} depicts predicted fractions of sight lines as a function of column densities of the observed weak Mg{\sc ii} absorbers at various redshifts by \citet{Rigby2002, Charlton2003, Ding2005, Misawa2008, Narayanan2008}, based on an assumption that they cover {$\sim5-30\%$ of a halo. The total covering fraction of weak Mg{\sc ii} absorbers in L$^{*}$ galactic haloes is estimated to be $\sim30\%$ by \citet{Narayanan2008, Muzahid2017}. If a sightline goes through N dwarf satellite galaxies, the covering fraction in each dwarf halo would need to be approximately 0.3/N. We would need a similar fraction of dwarf satellite haloes to be covered with weak Mg{\sc ii} absorbers, if undetected, satellite dwarf galaxies are responsible for producing weak Mg{\sc ii} absorbers in a L$^{*}$ halo. }  

Weak Mg{\sc ii} absorbers with column densities greater than the observed minimum, $\sim10^{11}$ cm$^{-2}$ occupy about only  $f_{MgII}\sim5\%$ of the dwarf halo in our simulation. {However, this is a lower limit for the covering fraction because $38-58\%$ of SNII outflow gas leaves the box by $t=40-300$~Myr.}
Boosting the metallicities of all the gas to 1 Z$_{\odot}$ (see \textit{dashed} lines in Figure~\ref{fig:coverf}) raises the fractions of sight lines with $N_{MgII}\gtrsim10^{11}$ cm$^{-2}$ to $f_{MgII}\sim30\%$, but there is still a deficiency of Mg{\sc ii} clouds with higher column densities, $\gtrsim10^{12}$ cm$^{-2}$. Most observed weak Mg{\sc ii} absorbers have column densities $\gtrsim10^{12}$ cm$^{-2}$.
As we argued in the previous section, repeated bursts of star formation will likely create more clumps and filaments, like the brightest structures in Figure~\ref{fig:CIVOVI}, through cycles of \textit{phase 1} and \textit{phase 2} formation. 
Then, a larger fraction of the dwarf galaxy halo may be covered with {moderately} dense Mg{\sc ii} absorbers. {However, the formation of denser, high column density, weak Mg{\sc ii} clouds may require other mechanisms that involve more gas and more metals with stronger shocks, as the shell density scales like the square of the Mach number in the isothermal shocks expected, so more powerful outflows may be responsible for the higher column density Mg{\sc ii} absorbers. In addition, interaction of outflows with cosmological infall will likely produce stronger shocks, so possibly denser clouds. Note we have a static background in our simulations. However, cooling is strongly limited by numerical resolution, so gas behind isothermal shocks will not cool to the theoretical density (n$\it{M}^{2}$) even with 0.2~pc resolution \citet{Fujita2009}. }

{We can estimate the number density of weak Mg{\sc ii} absorbers per unit comoving path length to be $dN_{MgII}/dX\approx$0.21
assuming $f_{MgII}\sim5\%$ for $N_{MgII}\ge10^{12}$ cm$^{-2}$ when metallicity is boosted to Z=Z$_{\odot}$, 1.13 Mpc$^{-3}$ for halo comoving number density with M$_{halo}$~$\geq3\times10^{9}~M_{\odot}$ at z=2 \citep{Murray}, and $\pi$($17.5^{2}-2.5^{2}$)~kpc$^{2}$ for halo proper cross section. This yields a value comparable to $dN_{MgII}/dX=0.33$ at $1.4<$z$<2.4$ found by \citet{Narayanan2008} and $dN_{MgII}/dX=0.41$ at $<$z$>$=2.34 by \citet{Codoreanu2018}. Likewise, the number density of high ionization clouds (C{\sc iv} and O{\sc vi}) per unit comoving path length is estimated to be $dN_{CIV}/dX\approx dN_{OVI}/dX\approx2$ with $f_{CIV}=f_{OVI}\sim50\%$. {As a reference, it is $dN_{CIV}/dX\approx9$ at z~$\sim$3 based on Figure 6 of \citet{DOdorico2016}, which includes all CIV systems along a line of sight, not necessarily only those confined to the CGM of galaxies. The covering fractions of C{\sc iv} and O{\sc vi} ions are measured to be 0.3-0.8 at impact parameters $\lesssim1$ pMpc around star-forming galaxies at z$\sim$2.4 \citep{Turner2014}.} It is interesting to note that the comoving Mg{\sc ii} mass density seems to increase nearly a factor of 10 from $<$z$>$=2.34 to $<$z$>$=4.77 \citep{Codoreanu2018} with a large number of weak Mg{\sc ii} absorbers even up to z$\sim$7 \citep{Bosman2017}. This high incidence of Mg{\sc ii} absorbers suggests that they are associated with dwarf galaxies, including smaller, numerous galaxies during the epoch of reionization, and the presence of the abundant weak Mg{\sc ii} absorbers must be explained without more powerful outflows from larger galaxies.}

We assess this as follows:
1) a SNII driven outflow is launched from a star cluster every 100 Myr, the time by which gas flows back to the central source region in our simulation, and  it takes 50 Myr for a SNII driven outflow with v=200-400 km s$^{-1}$ to reach the shocked enriched gas from previous outflows (region b). 2) SNIa drive a superbubble and an outflow after SNII stop in 50 Myr (we choose 50 instead of 40 Myr for simplicity), and it takes 100 Myr for a SNIa driven outflow to reach region (b) based on our simulation result. 3) repeated bursts last for 1 Gyr. 4) interaction from a newly launched outflow produces weak Mg{\sc ii} absorbers that cover 3-6$\%$ of our dwarf halo and those weak Mg{\sc ii} absorbers survive for at least 150 Myr based on our simulation result. Then, we estimate that the covering fraction of dwarf halos by weak Mg{\sc ii} absorbers will be 12-24$\%$. This number should go up once the CGM is more metal enriched, because the covering fraction of 3-6$\%$ is computed when metallicities of absorbers are Z=0.1-0.2 Z$_{\odot}$. We hope to test this hypothesis with our future global simulation in a larger box with repeated bursts in time and place. 
\begin{figure*}[!hbt]\centering
\includegraphics[width=0.8\textwidth, scale=0.8]{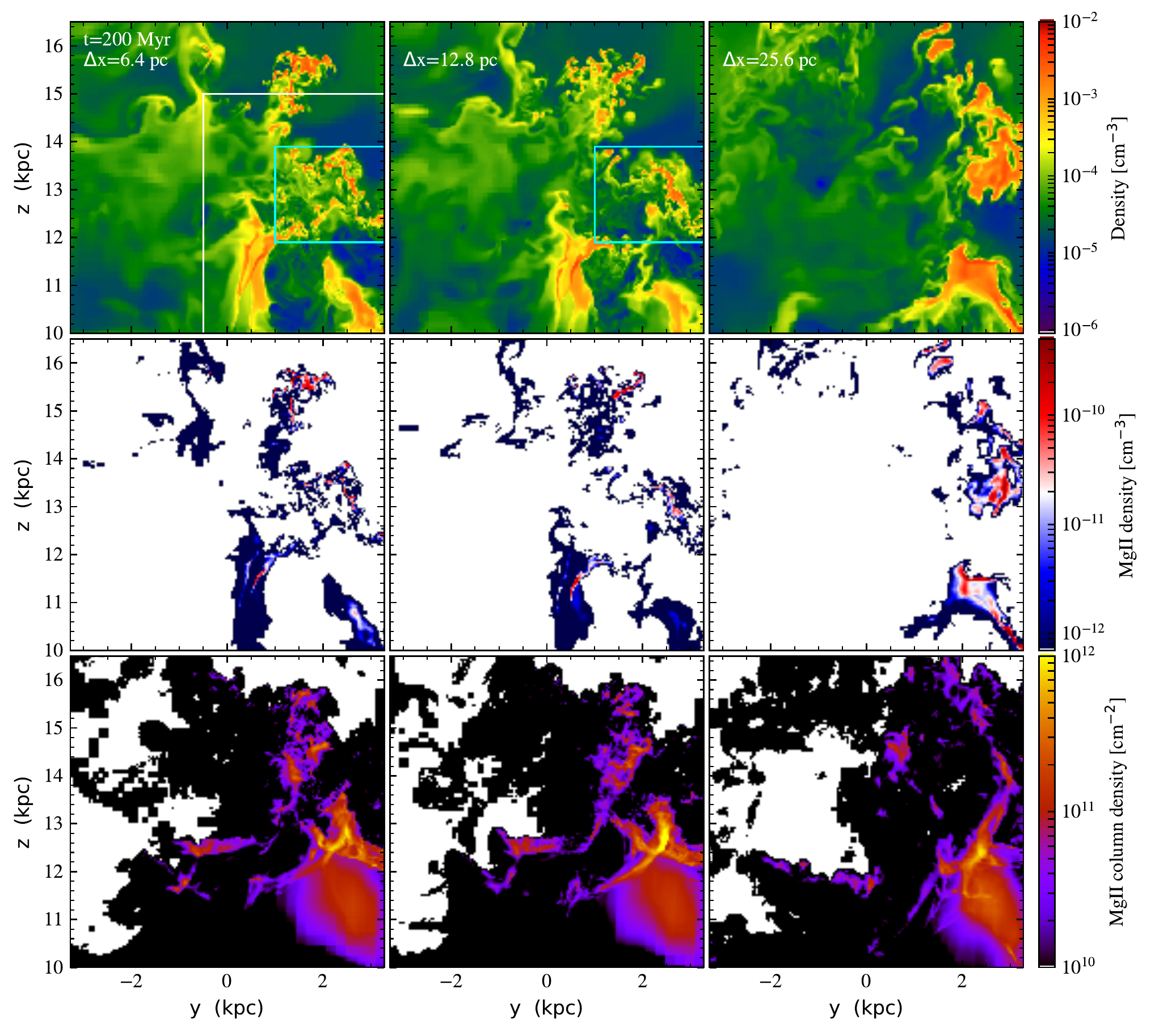}
    \caption{Sliced density (\textit{top}) and Mg{\sc ii} density (\textit{middle}) distributions at x=+2.4 kpc from the disk center and projected Mg{\sc ii} distributions (\textit{bottom}) along the $x$-axis, all in the y-z plane, at \textit{phase 1} (t=200 Myr), resolved with highest resolutions of $6.4$ in {[$\Delta$x,$\Delta$y,$\Delta$z]=[(-0.5 kpc, 3.28 kpc), (-0.5 kpc, 3.28 kpc), (10 kpc, 15 kpc)] (\textit{white rectangle}), 12.8 (our \textit{standard} simulation), and 25.6 pc (\textit{from left to right}). Regions enclosed in \textit{cyan} rectangles are shown in Figure~\ref{fig:res}. }
\label{fig:res9}}
\end{figure*}

\begin{figure*}[!hbt]\centering
\includegraphics[width=0.81\textwidth, scale=0.6]{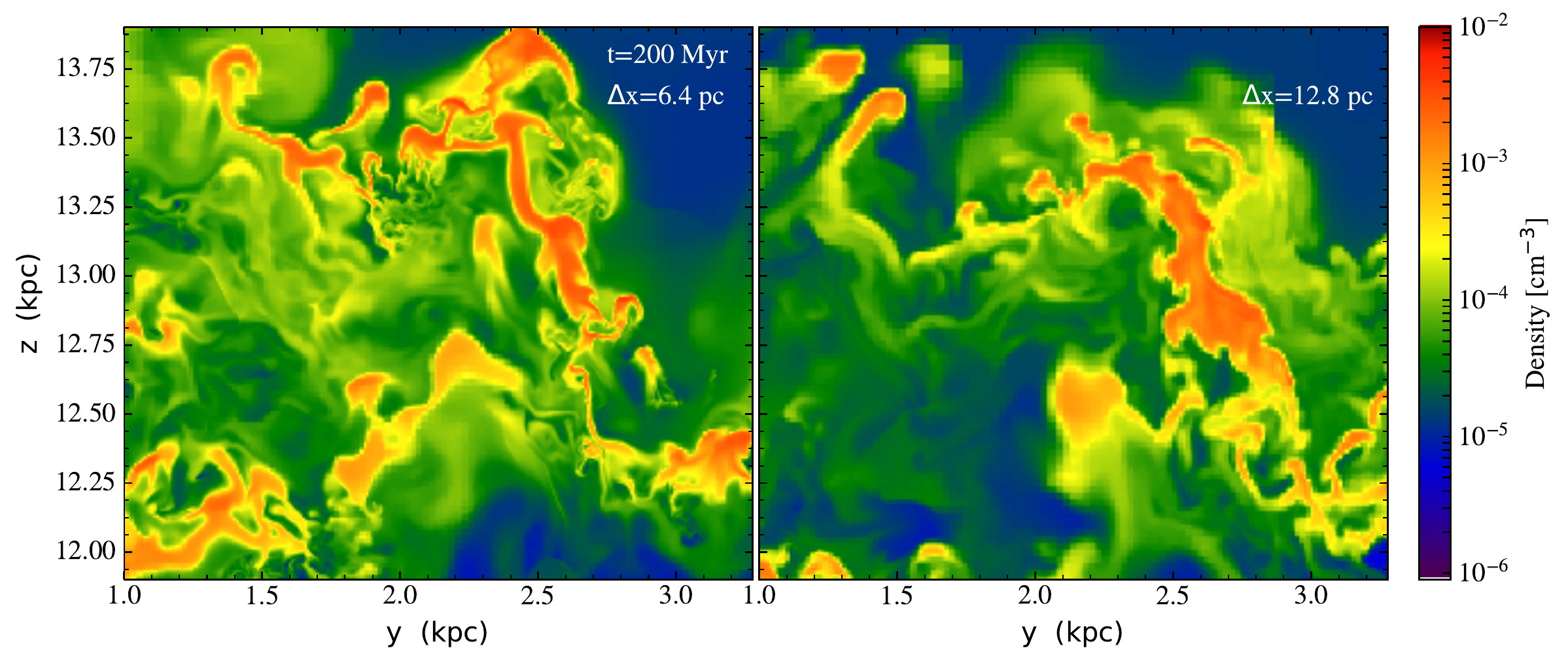}
\caption{The same as \textit{top} figures in Figure~\ref{fig:res9}, but   showing only regions enclosed in \textit{cyan} rectangles for 6.4 (\textit{left}) and 12.8 (\textit{right}) resolutions. 
\label{fig:res}}
\end{figure*}

\section{Resolution Study} \label{sec:resolution}

Our \textit{standard} simulation employs a highest resolution of
12.8~pc with four refinement levels, thus resolves $\sim100$ pc
structures
for our purposes. {We base the estimate of roughly eight cells being required
to minimally resolve structures on two arguments. First, the numerical
dissipation range for
supersonic turbulence computed with Enzo extends over almost an order
of magnitude \citep[e.g.][Figure 5]{kritsuk2007}, similar to most other grid
codes \citep{kitsionas2009}. Second, modeling of a
cloud in a supersonic flow shows that a radius of six zones using a second-order
method is insufficient to capture fragmentation by instabilities
\citep[Figure 4]{mac-low1994a}.}

To study the extent to which the production of clumps and filaments as well as their sub-structures and fragmentation are dependent on numerical resolution, we ran the same simulation with 3 refinement levels (\textit{low-res} simulation), and by applying 5 refinement levels in a region where the largest filaments form at [$\Delta$x,$\Delta$y,$\Delta$z]=[(-0.5 kpc, 3.28 kpc), (-0.5 kpc, 3.28 kpc), (10 kpc, 15 kpc)] (\textit{high-res zoom} simulation). We only ran the \textit{high-res zoom} simulation up to t=200 Myr. 

Figure~\ref{fig:res9} shows \textit{phase 1} formation of filaments and clumps computed with the three different resolutions. Figure~\ref{fig:res9} compares the degrees of fragmentation in \textit{high-res zoom} and our \textit{standard} simulations. In the \textit{high-res zoom} simulation, gas fragments into thinner filaments and smaller clouds compared with our \textit{standard} simulation. The smallest structures are resolved across $\sim$8 cells, so they are $\sim$50 pc in the \textit{high-res zoom} simulation compared with $\sim$100 pc in our \textit{standard} simulation. These filaments and clumps will further fragment into smaller pieces with higher resolution. Gas structures seem drastically different in the \textit{low-res} simulation, with much larger clouds compared with the higher resolution runs. 

Despite the differences in fragmentation seen in simulations with different resolutions, there is no significant difference in projected Mg{\sc ii} distributions (\textit{bottom} figures). We see no change in the fraction of weak Mg{\sc ii} absorbers with high column densities, and the covering fractions of weak Mg{\sc ii} absorbers as well as C{\sc iv} and O{\sc vi} absorbers remain practically the same. {We will add ion versus H{\sc i} column density distributions with metallicities (Figure~\ref{fig:scatterg}) and ion covering fractions as functions of column densities (Figure~\ref{fig:coverfg}) for the \textit{low-res} simulation in the appendix}. 

We conclude that resolution has a small effect on the {projected distribution} of dense clumps and filaments, {and this in turn warns us to be very careful when we interpret observations \citep{Peeples2019}}. There is a possibility that, at much higher resolution filaments and fragments will further "shatter" into $\sim$pc sized cloudlets that agree with some of the observed Mg{\sc ii} absorbers \citep{Gronke2018, 2018MNRAS.473.5407M,2020MNRAS.494L..27G,BegelmanFabian1990},
{but to test this possibility requires 1/10th pc resolution in a galactic-scale simulation which is not feasible at the moment.} {Opposing fragmentation is possible cloud coalescence, which may be numerically challenging to reproduce in 3D simulations \citep{2019ApJ...876L...3W}.}

\section{Summary} \label{sec:summary}
In this paper, we use hydrodynamical simulations of galactic outflows to explore the production of 
weak Mg{\sc ii} absorbers and C{\sc iv} and O{\sc vi} absorbers in the CGM of a
dwarf satellite galaxy with a halo mass of $5\times10^{9}$~M$_{\odot}$ at $z=2$, expected to be hosted by a larger $L^{*}$ halo. With our standard numerical resolution of 12.8~pc, we model the formation of superbubbles and outflows from a galactic disk assuming a single instantaneous starburst in a simulation box with dimensions (6.5536, 6.5536, 32.768)~kpc, and study the interaction and cooling of metal enriched outflowing gases. Although we ran the simulations only for a duration of $\sim$300 Myr, as most metal enriched gas leaves the simulation box by this time, our results highlight the possibility of dwarf galactic outflows producing transient, but continuously generated Mg{\sc ii} clouds as well as larger C{\sc iv} and O{\sc vi} clouds in sub-LLS and Ly$\alpha$ forest environments.

Our main findings are:
\begin{itemize}
	\item Thin, filamentary, weak Mg{\sc ii} absorbers are produced in
two stages: \begin{itemize}
	\item \textit{phase 1} shocked SNII enriched gas loses energy and descends toward expanding SNII enriched gas and is shocked, cools, and fragments.
	\item \textit{phase 2} SNIa driven outflow gas shocks the SNII enriched gas as well as \textit{phase 1} shells, which then cool and fragment. 
	\end{itemize}
	The width of the filaments and fragments are $\lesssim~100$~pc with our standard numerical resolution. A single Mg{\sc ii} cloud survives for $\sim$~60 Myr, but we suggest Mg{\sc ii} absorbers will continuously be produced through cycles of \textit{phase 1} and \textit{phase 2} formation for $>150$ Myr by repeated bursts of star formation. 
	\item C{\sc iv} absorbers are produced in expanding SNII enriched gas (region a) and shocked SNII enriched gas (region b). C{\sc iv} absobers in region (a) extend over 1--4~kpc and C{\sc iv} absorbers in region (b) are smaller, 500~pc--1~kpc, but they are both cool and photoionized. The smaller C{\sc iv} absorbers originate from the same clouds that produce weak Mg{\sc ii} absorbers, and they surround the dense Mg{\sc ii} clouds. As the clouds get destroyed and mixed with the surrounding gas, Mg{\sc ii} absorbers disappear first, but C{\sc iv} absorbers survive for another 20--30 Myr. 
	\item O{\sc vi} absobers are also produced in expanding SNII enriched gas in region (a) and shocked SNII enriched gas in region (b). O{\sc vi} absorbers in region (a) originate from the same cool clouds that produce C{\sc iv} absorbers, but O{\sc vi} absorbers in region (b) are not coincident with Mg{\sc ii} absorbers or C{\sc iv} absorbers. Their sizes are over $\gtrsim$1~kpc.  
	\item {C{\sc iv} absorbers and most O{\sc vi} absorbers  are cool, photoionized clouds while O{\sc vi} absorbers arising in turbulent mixing layers in region (b) are hotter and collisionally ionized. Photoinization dominates in sub-LLS and Ly$\alpha$ environments at intermediate redshft .}
	\item The metallicities of Mg{\sc ii}, C{\sc iv}, and O{\sc vi} absorbers are Z=0.1--0.2~Z$\odot$ by t=$\sim$200--300 Myr, with only one cycle of \textit{phase 1} and \textit{phase 2} formation in a dwarf disk and halo with a low initial metallicity, Z=0.001~Z$\odot$.  We speculate that the clouds forming in shocked outflow gas (region b) {will be progressively enriched with more metals when bursts of star formation are repeated.}  
	\item {A lower limit for the covering fraction of weak Mg{\sc ii} absorbers in our dwarf halo is 3--6$\%$ because we compute only one cycle of \textit{phase 1} and \textit{phase 2} formation and more than half the metal enriched gas leaves the simulation box early.} {To reproduce the observed estimate for the covering fraction in a L$^{*}$ halo (30$\%$) with outflows from such galaxies alone, sightlines must go through haloes of multiple dwarf satellite galaxies.  We also speculate that the covering fraction in a single dwarf halo will be boosted with repeated bursts with many cycles of \textit{phase 1} and \textit{phase 2} formation in a large simulation box that covers the entire halo.}

\end{itemize}
{There are two major problems in our current simulations: 1) a deficiency of weak Mg{\sc ii} absorbers with high column density, $\gtrsim10^{12}$ cm$^{-2}$ and 2) low metallicity of weak Mg{\sc ii} absorbers. 

1) The formation of denser, high column, weak Mg{\sc ii} absorbers may require stronger shocks with more powerful outflows and/or outflows interacting with dynamic infall. Repeated outflows shocking the clumps and filaments formed by previous outflows may also produce denser Mg{\sc ii} clouds. However, cooling behind shocks is limited by numerical resolution. {Our resolution study shows that the sizes of filaments and fragments decrease by a factor of two with a resolution twice as high, however, the projected properties are insensitive to changes in resolution that is $\gg$~pc.} We may need less than a pc scale resolution to address this problem. 

2) The metallicity, less than solar, of our Mg{\sc ii} absorbers is the result of our assumption of a single instantaneous starburst and the limited duration of our simulations ($\sim300$ Myr) neglecting the SNIa metal contribution. Starting with a higher initial metallicity for our dwarf disk and halo gas will also alleviate the problem. 

This paper nonetheless highlights the possibility of galactic outflows from invisible, dwarf satellite galaxies to produce highly enriched, multiphase gas.
We hope to address the remaining problems with our next, more global simulations. }

\acknowledgments

This work was supported by the Grants-in-Aid for Basic Research by the Ministry of Education, Science and Culture of Japan, Grant Number 19K03911. We acknowledge use of the Cray XC50 at the Center for
Computational Astrophysics (CfCA) of the National Astronomical
Observatory of Japan (NAOJ). J.C.C. acknowledges support by the National Science Foundation under grant No. AST-1517816.
AM acknowledges support from UK Science and Technology Facilities Council, Consolidated Grant ST/R000972/1.
M-MML was partly supported by US NSF grant AST18-15461.
Computations described in this work were performed using the publicly available \texttt{Enzo} code (http://enzo-project.org), which is
the product of a collaborative effort of many independent scientists from
numerous institutions around the world.  Their commitment to open science
has helped make this work possible.
\vspace{5mm}
\facilities{CfCA(NAOJ)}


\software{Enzo \citep{Bryanetal2014}, yt \citep{Turk2011}
Trident \citep{Hummels2017},  
           SYGMA \citep{RItter2018}}
           
\clearpage
\appendix
\begin{figure*}[!ht]
{\centering
\includegraphics[width=0.9\textwidth, scale=0.6]{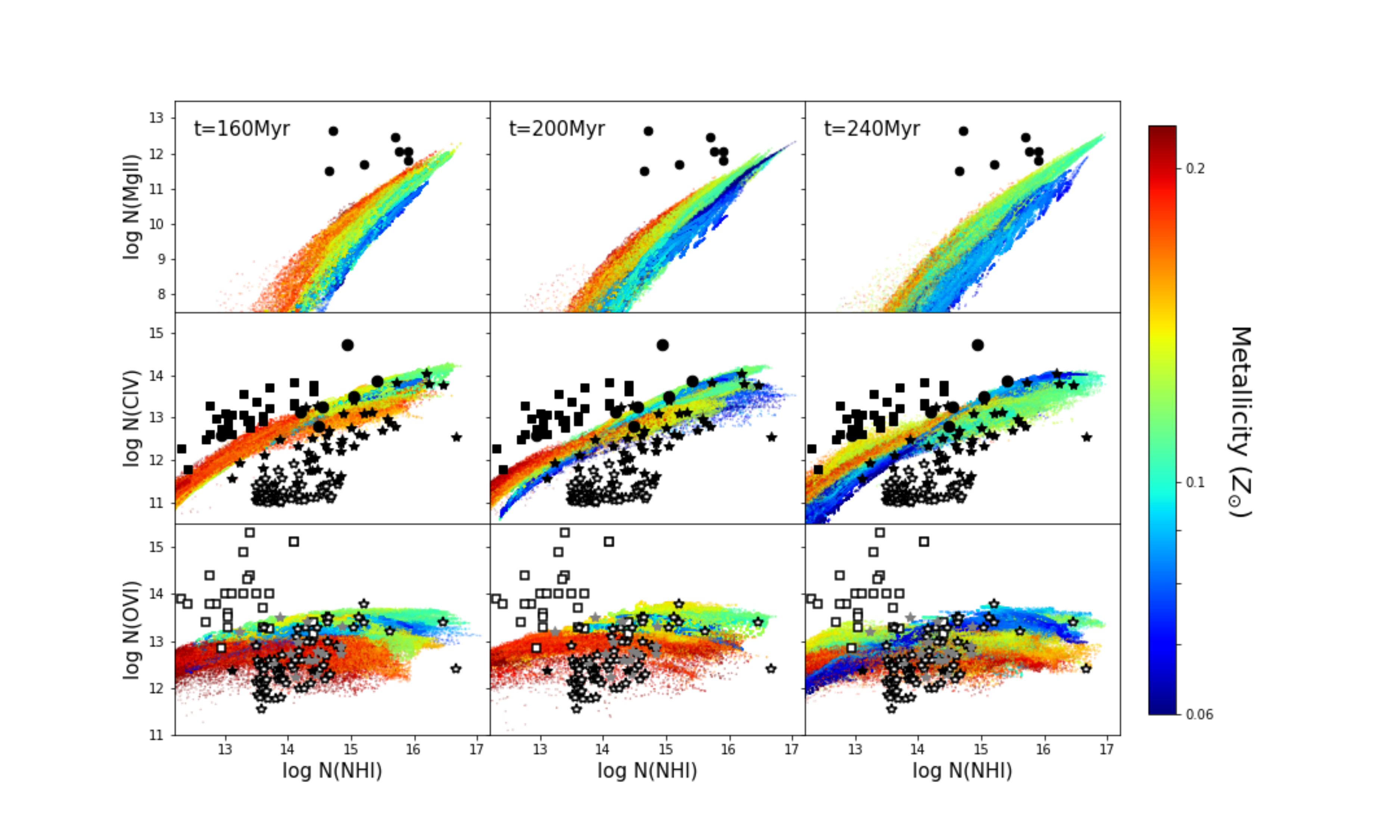}
\figcaption{Same as Figure~\ref{fig:scatter} for \textit{low-res}
  simulation (25.6 pc) 
\label{fig:scatterg}} 
}
\end{figure*}

\begin{figure*}[!ht]
{\centering
\includegraphics[width=0.65\textwidth, scale=0.6]{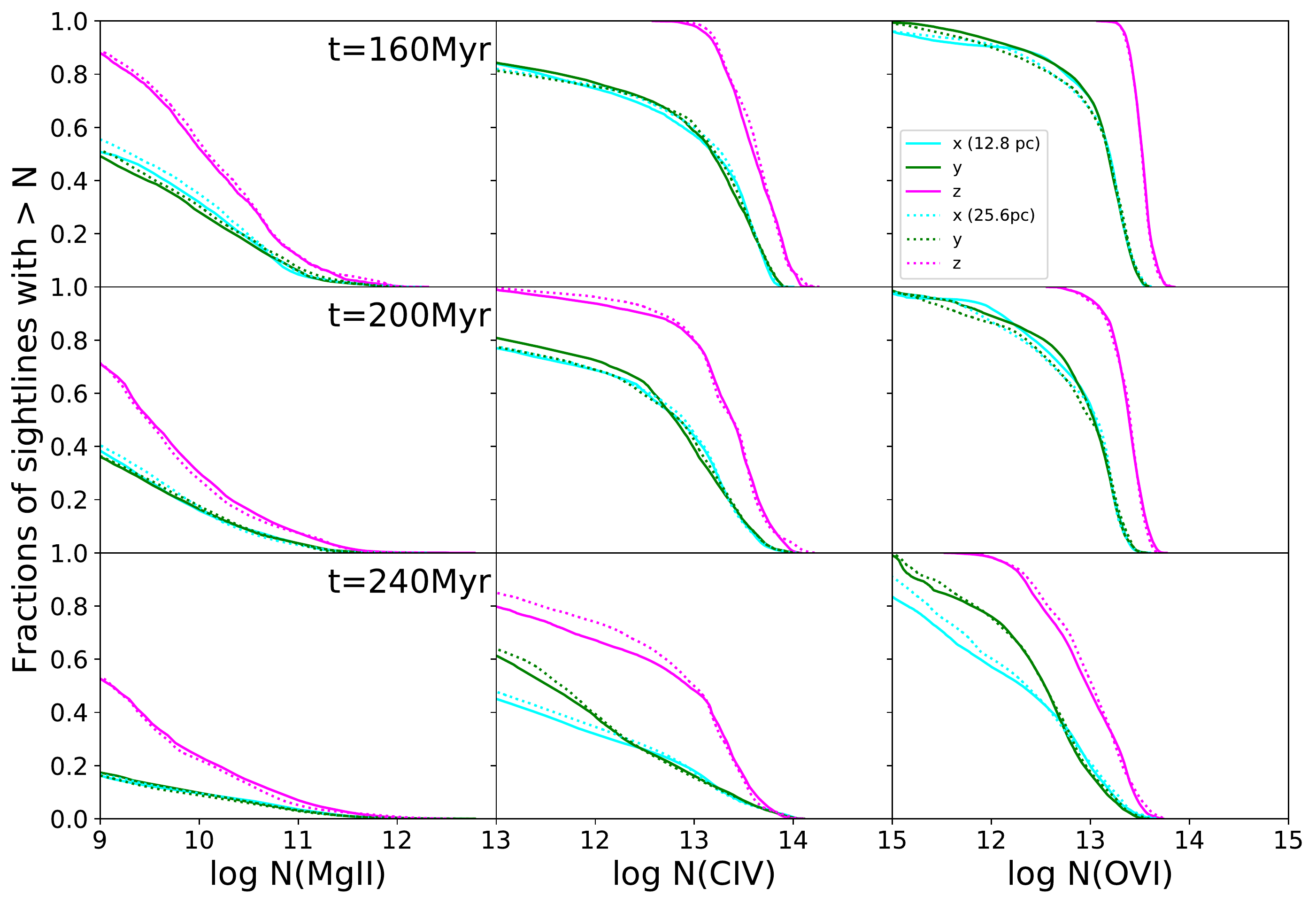}
\figcaption{Same as Figure~\ref{fig:coverf} but comparing the results in 
 in our \textit{standard} simulations (12.8 pc \textit{solid}) and \textit{low-res}
  simulation (25.6 pc \textit{dashed}). \label{fig:coverfg}}
  }
\end{figure*}

\begin{figure*}[!ht]
{\centering
\includegraphics[width=0.9\textwidth, scale=0.6]{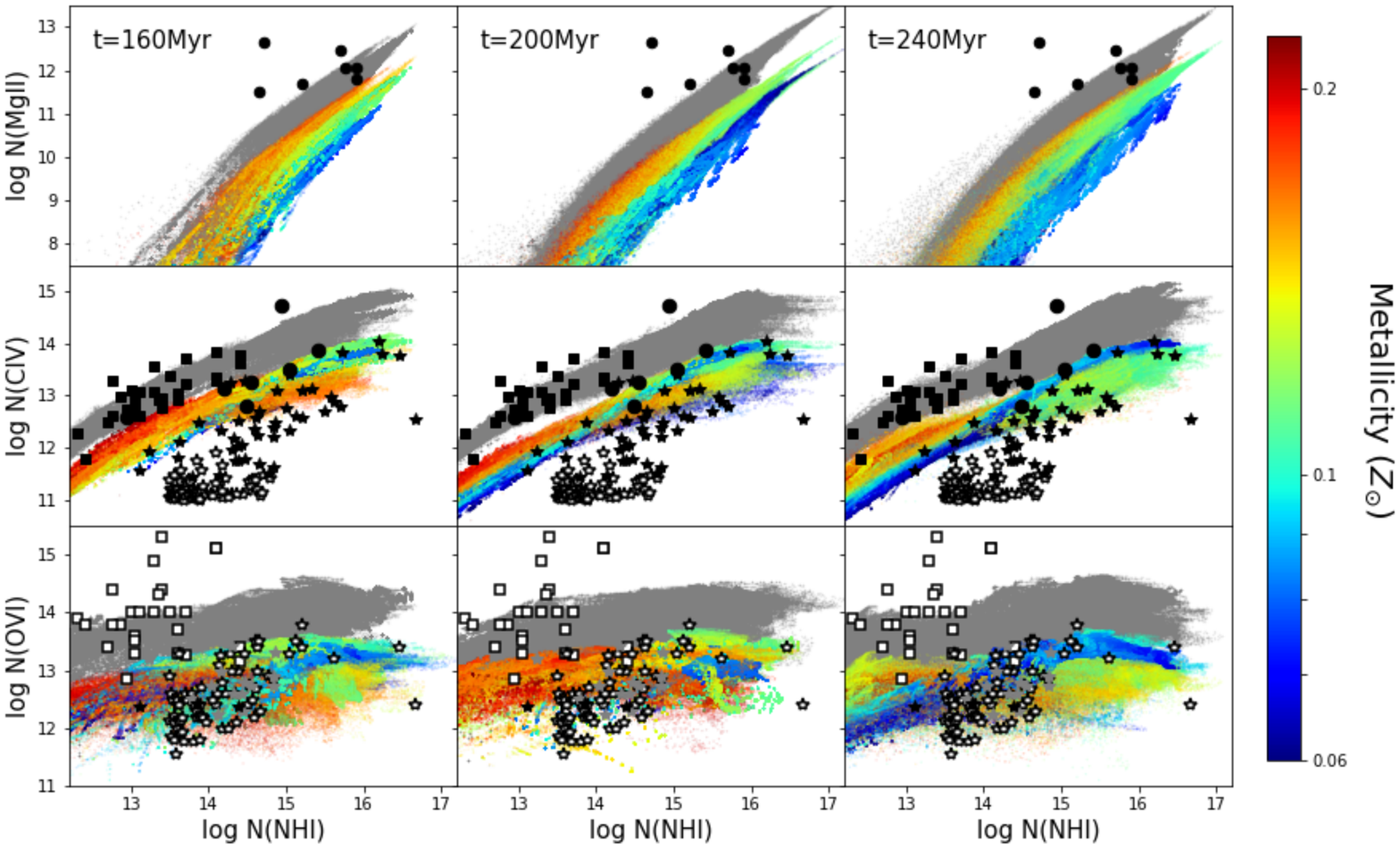}
\figcaption{Same as Figure~\ref{fig:scatter} but without overcooled gas (n$_{H}\leq 10^{-4}$ cm$^{-3}$, T$< 10^{4}$ K) in our \textit{standard}
  simulation. The overcooled, low-density gas is metal-enriched outflow gas in region a. With or without it, there is very little change for Mg{\sc ii}  and C{\sc iv}  distributions, while there is a marginal difference in the distribution of higher metallicity O{\sc vi}  systems. 
\label{fig:scatterOC}} 
}
\end{figure*}

\begin{figure*}[!ht]
{\centering
\includegraphics[width=0.6\textwidth, scale=0.6]{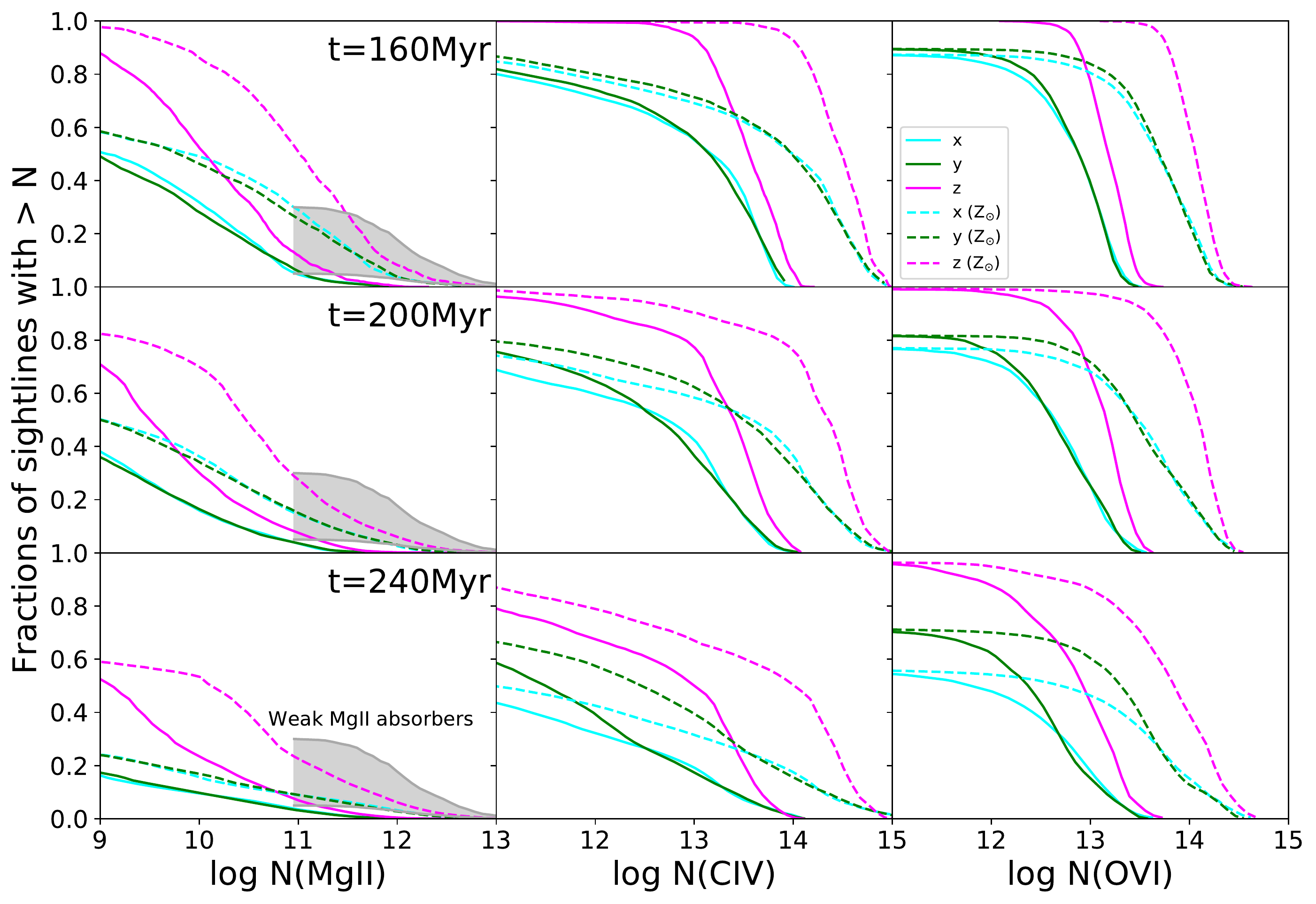}
\figcaption{Same as Figure~\ref{fig:scatter} {{but without overcooled gas (n$_{H}\leq 10^{-4}$ cm$^{-3}$, T$< 10^{4}$ K) in our \textit{standard}
  simulation. With or without it, there is no noticeable difference in the covering fractions of Mg{\sc ii}  and C{\sc iv} systems, but there is a slight decrease in the covering fraction of O{\sc vi} systems at lower column density.}}
\label{fig:coveringfractionOC}} 
}
\end{figure*}

\clearpage
%



\bibliography{draft2}{}
\bibliographystyle{aasjournal}



\end{document}